\documentclass[11pt]{article}

\usepackage{amsfonts}
\usepackage{amsmath}
\usepackage{esint}
\usepackage[margin=1in]{geometry}
\usepackage{graphicx}
\usepackage{epstopdf}
\usepackage{amssymb}
\usepackage{cite}
\usepackage{hyperref}
\usepackage{wrapfig}
\usepackage{subcaption}
\usepackage{float}
\usepackage{caption}
\usepackage{color}
\usepackage{array}
\usepackage{url}

\usepackage{multicol}

\title{Modeling Microglia Activation and Inflammation-Based Neuroprotectant Strategies During Ischemic Stroke} 
\author{Sara Amato, Andrea Arnold*}
\date{}

\begin{document}
\maketitle

\small 

\centerline{Department of Mathematical Sciences, Worcester Polytechnic Institute, Worcester, MA, USA}
\vspace{.2cm}

\centerline{* corresponding author: anarnold@wpi.edu}

\normalsize

\bigskip

\begin{abstract}

Neural inflammation immediately follows the onset of ischemic stroke. During this process, microglial cells can be activated into two different phenotypes: the M1 phenotype, which can worsen brain injury by producing pro-inflammatory cytokines; or the M2 phenotype, which can aid in long term recovery by producing anti-inflammatory cytokines. In this study, we formulate a nonlinear system of differential equations to model the activation of microglia post-ischemic stroke, which includes bidirectional switching between the microglia phenotypes, as well as the interactions between these cells and the cytokines that they produce. Further, we explore neuroprotectant-based modeling strategies to suppress the activation of the detrimental M1 phenotype, while promoting activation of the beneficial M2 phenotype. Through use of global sensitivity techniques, we analyze the effects of the model parameters on the ratio of M1 to M2 microglia and the total number of activated microglial cells in the system over time. Results demonstrate the significance of bidirectional microglia phenotype switching on the ratio of M1 to M2 microglia, in both the absence and presence of neuroprotectant terms. Simulations further suggest that early inhibition of M1 activation and support of M2 activation leads to a decreased minimum ratio of M1 to M2 microglia and allows for a larger number of M2 than M1 cells for a longer time period. \\

\end{abstract}

\section{Introduction}
Stroke is the second leading cause of death worldwide and the fifth leading cause of death in the United States, as well as a leading cause of disability \cite{Johnson2016, WHO, ASA, CDC}. Ischemic strokes account for $87\%$ of all strokes and are caused by a blockage in a blood vessel due to thrombosis or embolism, resulting in oxygen deprivation of the brain \cite{ASA}. During an ischemic stroke, cell death and damage occur in the affected brain area called the infarcted core \cite{Newton2016, Taylor2013, Yenari2010}. Immediately following the onset of ischemia, the body naturally responds with inflammation, which can both worsen brain injury and help in long term recovery.

The goal of this study is to develop a mathematical model of the neuroinflammatory process during ischemic stroke to analyze both the beneficial and detrimental effects of inflammation. In particular, we introduce a new coupled system of nonlinear differential equations to model the dynamic interactions between microglia and cytokines, two of the main components involved in neuroinflammation following stroke onset. Neuroinflammation begins with the activation of microglia, a type of neuroglia residing in the central nervous system \cite{Anderson2015, Hu2012, Orihuela2016, Russo2010, Taylor2013, Yenari2010}. This activation peaks around two to three days after stroke and persists for several weeks \cite{Lee2014,  Guruswamy2017}. Microglia maintain homeostasis of the brain by continuously monitoring their surrounding environment and responding to pathological signals released by neighboring brain cells \cite{Byrne2014, Yenari2010, Boche2013}. Based on their type, activated microglia produce either anti-inflammatory cytokines or pro-inflammatory cytokines, thereby causing both beneficial and detrimental effects on the brain post-ischemia.

Microglia activation is characterized by two phenotypes: M1 and M2. The M1 phenotype (classical activation) is characterized by the secretion of pro-inflammatory cytokines, which can exacerbate the inflammatory response and lead to further brain damage. Pro-inflammatory cytokines include tumor necrosis factor alpha (TNF-$\alpha$), interleukin 1 beta (IL-$1\beta$), nitric oxide, interleukin 6 (IL-$6$), and interleukin 12 (IL-$12$) \cite{Taylor2013, Orihuela2016, Hu2015, Tang2016, Hao2016, Nakagawa2014, Cherry2014, Shao2013}. Microglia can also be activated into the M2 phenotype (alternative activiation) and perform crucial roles in limiting inflammation by releasing anti-inflammatory cytokines, including interleukin 4 (IL-$4$), interleukin 10 (IL-$10$), and transforming growth factor beta (TGF-$\beta$) \cite{Taylor2013, Orihuela2016, Tang2016, Hao2016, Nakagawa2014, Ledeboer2000, Liu2016, Hu2015, Shao2013}.

M2 microglia have been shown to dominate at the early stages of inflammation, whereas M1 microglia activate more slowly and then become the dominant phenotype for the remainder of the neuroinflammatory process \cite{Hu2012, Tang2016}. Classical and alternative microglia activation is positively influenced by the presence of pro-inflammatory cytokines and anti-inflammatory cytokines, respectively \cite{Orihuela2016, Shao2013, Nakagawa2014, Vaughan2018,Liu2016}. Experimental studies have shown that anti-inflammatory cytokines inhibit microglia activation to the M1 phenotype and, on the other hand, pro-inflammatory cytokines inhibit microglia activation to the M2 phenotype \cite{Shao2013, Orihuela2016, Hu2015, Taylor2013, Tang2016}. Further, experimental studies have also shown that microglia may switch phenotypes from M1 to M2 and vice versa \cite{Zhao2017, Hu2015, Tanaka2015, Orihuela2016, Qin2019, Nakagawa2014, Guruswamy2017}. The switching from the M2 to M1 phenotype has been cited as an area in need of further research \cite{Boche2013}. Mathematical models for applications other than ischemic stroke have considered interactions between microglia phenotypes but have not included the possibility of switching from the M2 to the M1 phenotype \cite{Hao2016, Wang2012, Vaughan2018}.

In this study, we develop a four-compartment model of microglia and cytokine interaction, which includes both the M1 and M2 phenotypes, pro-inflammatory and anti-inflammatory cytokines, and bidirectional phenotype switching between M1 and M2 microglia. Previous mathematical models studying intracellular processes of ischemic stroke inflammation have included terms representing general microglia activation, leukocytes, astrocytes, and neurons \cite{Russo2010, Dronne2005, Orlowski2011, Newton2016, Rayz2008, Lelekov2009}. However, these models do not consider the two microglia phenotypes or phenotype switching, which we include in this work to analyze both the beneficial and deleterious effects of microglia activation post ischemic stroke. We also study specifically the effects of M2 to M1 phenotype switching, which may lead to increased cell damage by bolstered production of pro-inflammatory substances in the brain.  

Mathematical models of neuroinflammation have included the two microglia phenotypes for applications other than stroke, including traumatic brain injury (TBI), amyotrophic lateral sclerosis (ALS), hemorrhagic shock, and Alzheimer's disease \cite{Vaughan2018, Shao2013, Hao2016}. The models for TBI in \cite{Vaughan2018} and for Alzheimer's disease in \cite{Hao2016} include switching from M1 to M2 but do not include switching from the M2 to the M1 phenotype. The model for ALS presented in \cite{Shao2013} includes bidirectional switching between microglia phenotypes; however, to the authors' knowledge, there are no current models of neuroinflammation during ischemic stroke which include bidirectional microglia phenotype switching. Further studies have explored the interactions between cytokines in general neural inflammation \cite{Anderson2015, Torres2009} but have not included the interactions of microglia producing these substances. Multiple studies have also explored inflammation with macrophages, which behave in a similar manner to microglia, in applications such as myocardial infarction \cite{Malek2015, Wang2012}. 

Despite the widespread impact of ischemic stroke, there are currently only two clinical treatment strategies for clot removal. Tissue plasminogen activator (tPA)-induced thrombolysis is the only FDA-approved medication to restore blood flow in the brain following ischemia. During this treatment, tPA is intravenously administered to break up the clot within the blood vessel that is causing the ischemic stroke. This strategy is limited to a small subset of stroke patients due to its short treatment window \cite{Guruswamy2017, Kent2006, Piebalgs2018, Gu2019, Minnerup2012}. An alternative to thrombolysis drug treatment is thrombectomy, a surgical procedure during which a clot retrieval device is used to mechanically remove the blood clot causing the ischemic stroke. Mathematical models for both thrombolysis drug treatment and thrombectomy have been developed, including: compartment models to evaluate the effects of tPA dose on the effectiveness of treatment \cite{Piebalgs2018, Gu2019}; predictive models to identify subsets of patients who would be eligible for thrombolysis \cite{Kent2006}; and a model of clot removal for mechanical thrombectomy \cite{Romero2011}. Both of these treatment strategies increase the risk for hemorrhage post ischemic stroke \cite{Motto1999, Shoamanesh2013, Neuberger2019}. 

A potential new therapeutic strategy may be to target the distinct microglia phenotypes and promote M2 activation while simultaneously suppressing M1 activation \cite{Lee2014, Guruswamy2017, Ginsberg2008, Zhao2017, Yenari2010}. Recent experimental studies have explored the use of neuroprotective substances such as BHDPC, curcumin, miR-124, salidroside (SLDS), glycine, and celastrol to achieve this aim. An in vitro study showed that BHDPC, a novel neuroprotectant, was able to promote M2 phenotype polarization \cite{Li2018}. In a follow-up study, BHDPC was shown to reduce the amount of M1 microglial cells and enhance the amount of M2 microglia in middle cerebral artery occlusion-induced ischemic brain in mice after treatment with the drug \cite{Li2019}. Curcumin was shown to promote M2 microglia activation and inhibit pro-inflammatory responses both {in vitro} and {in vivo} in mice \cite{Liu2017}. Injection with the microRNA miR-124 was shown to decrease of ratio of M1 to M2 microglia in a mouse model \cite{Taj2016}. Intravenous SLDS injection decreased M1 microglial cells and increased M2 microglial cells post ischemic stroke in a mouse model \cite{Liu2018}. The amino acid glycine was shown to promote M2 microglial cells {in vitro} and {in vivo} in Sprague-Dawley rats \cite{Liu2019}. Celastrol was shown to decrease pro-inflammatory cytokines in several studies using rodent models \cite{Jiang2018, Li2012}.

Inspired by neuroprotectant strategies, we further modify the model proposed in this work to include time-varying terms aiding in the activation of M2 microglia and inhibiting the activation of M1 microglia. We analyze the effects of these neuroprotectant-based terms on the total number of activated microglia in the system, with specific interest in the ratio of M1 to M2 microglia, for different simulated treatment onset times. Further, by employing global sensitivity techniques, we analyze the effects of the model parameters on the ratio of M1 to M2 cells and the total activated microglia in both the absence and presence of the neuroprotectant terms. Results emphasize the significance of the microglia phenotype switching rates with respect to model sensitivity when considering the ratio of M1 to M2 microglia, while the number of resting microglia and the microglia activation and mortality rates are more significant when considering the total activated microglia.

The paper is organized as follows. Section \ref{sec: Model Description} describes the coupled system of nonlinear differential equations derived to model the interactions between the two microglia phenotypes and the pro- and anti-inflammatory cytokines. Section \ref{sec: Sensitivity Analysis} reviews two global sensitivity analysis techniques, Morris elementary effects and the Sobol method, and provides numerical results when these techniques are applied to the model derived in Section \ref{sec: Model Description}. Section \ref{sec: Modified Model} details the neuroprotectant-based terms added to the model to suppress M1 microglia production and bolster M2 microglia production. This section also provides computational simulations of the modified model when the neuroprotectant onset times are varied and sensitivity analysis of the modified model. Section \ref{sec: Discussion} features a discussion of the results and future work, and Section \ref{sec: Conclusion} gives a summary and conclusions of this work.

\section{Model Description} \label{sec: Model Description}
\begin{figure}[b!]
\centering
  \includegraphics[width=0.45\linewidth]{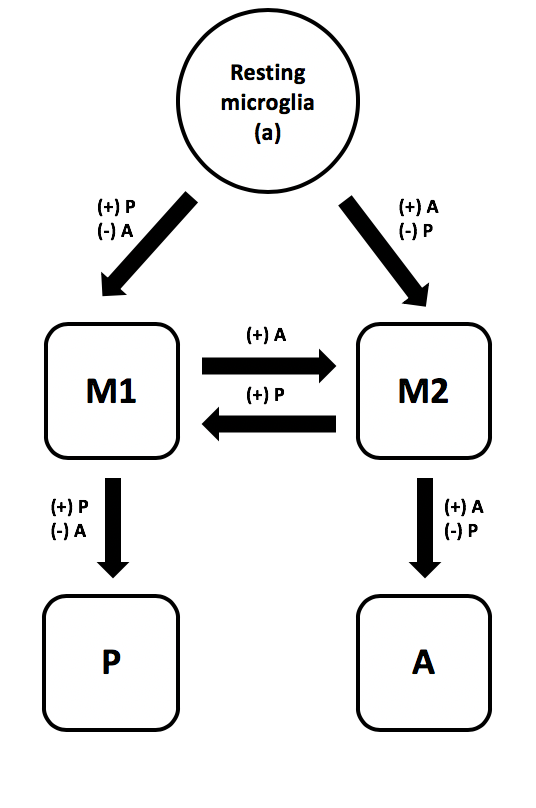}
  \caption{Schematic representation of microglia activation. After the onset of stroke, resting microglia are activated into the M1 or M2 phenotype. Activation to the M1 phenotype is positively influenced by the presence of pro-inflammatory cytokines ($P$) and negatively influenced by anti-inflammatory cytokines ($A$); vice versa for the M2 activation. M1 microglia release detrimental pro-inflammatory cytokines, and this production is positively influenced by the concentration of pro-inflammatory cytokines and negatively influenced by the concentration of anti-inflammatory cytokines. Conversely, M2 microglia produce beneficial anti-inflammatory cytokines, and this production is positively influenced by the concentration of anti-inflammatory cytokines and negatively influenced by the concentration of pro-inflammatory cytokines. Activated microglia may switch between the M1 and M2 phenotypes, with the M1 to M2 switching being positively influenced by the anti-inflammatory cytokines and the M2 to M1 switching being positively influenced by the concentration of pro-inflammatory cytokines. }
  \label{fig: Model}
\end{figure}

In this section we derive a simplified model of neural inflammation post-ischemic stroke, focusing on the interaction between the M1 and M2 microglia phenotypes and pro- and anti-inflammatory cytokines. We assume a constant source of resting microglia, which activates into the M1 or M2 phenotypes immediately following the onset of ischemic stroke. This activation is assumed to occur at a constant rate, influenced by the cytokine concentrations. Bidirectional switching can occur between the M1 and M2 microglia phenotypes. Pro-inflammatory ($P$) and anti-inflammatory ($A$) cytokines are produced by the M1 and M2 microglia, respectively, further influenced by the current concentrations of cytokines.  Figure~\ref{fig: Model} gives a schematic representation of the model.

\subsection{Microglia}
The equations describing the change in M1 and M2 microglial cells are as follows:
\begin{equation}\label{eq:M1}
\frac{dM1}{dt} = \underbrace{a \cdot k_{M1} \cdot H(P) \cdot \hat{H}(A)}_\text{M1 activation} - \underbrace{s_{M1 \rightarrow M2} \cdot H(A) \cdot M1}_\text{M1 to M2 switching} + \underbrace{s_{M2 \rightarrow M1} \cdot H(P)  \cdot M2}_\text{M2 to M1 switching} - \underbrace{\mu_{M1} \cdot M1}_\text{M1 mortality}  
\end{equation}
\begin{equation}\label{eq:M2}
\frac{dM2}{dt} = \underbrace{a \cdot k_{M2} \cdot H(A) \cdot \hat{H}(P)}_\text{M2 activation} + \underbrace{s_{M1 \rightarrow M2} \cdot H(A) \cdot M1}_\text{M1 to M2 switching} - \underbrace{s_{M2 \rightarrow M1} \cdot H(P)  \cdot M2}_\text{M2 to M1 switching} - \underbrace{\mu_{M2} \cdot M2}_\text{M2 mortality}  
\end{equation}
where $a$ is the number of resting microglia, $s_{M1 \rightarrow M2}$ and $s_{M2 \rightarrow M1}$ are constant parameters for microglia phenotype switching, and $k_{M1}$, $k_{M2}$, $\mu_{M1}$, and $\mu_{M2}$ are constant parameters for the activation and natural mortality of the microglial cells, respectively.  The Hill functions $H$ and $\hat{H}$ of the cytokines are of the form
\begin{equation} \label{eq:H}
H(x) = \frac{x^{n_x}}{x^{n_x} + K_{x}^{n_x}} 
\end{equation}
and
\begin{equation} \label{eq:Hhat}
\hat{H}(x) = \frac{K_x^{N_x}}{K_x^{N_x} + x_{x}^{N_x}}
\end{equation}
where $x$ is taken to be either $P$ or $A$ (representing the pro-inflammatory or anti-inflammatory cytokines, respectively), $n_x$ and $N_x$ are constant exponents which control the steepness of the curves, and $K_x$ is the half maximal concentration of the respective cytokine. Note that $H(x)$ has the form of an increasing sigmoidal curve, whereas $\hat{H}(x)$ is a decreasing sigmoidal curve. Similar Hill functions have been used in previous modeling of cytokines and cells \cite{Anderson2015, Vaughan2018, Wang2012, Hao2016, Shao2013, Malek2015}. The following subsections detail the model terms for microglia activation and phenotype switching. Table \ref{tab: ParameterTable} lists the descriptions, units, and nominal values of the corresponding model parameters.

\subsubsection*{Microglia activation}
Following the onset of ischemic stroke, resting microglia can become polarized to the M1 and M2 phenotypes \cite{Orihuela2016, Taylor2013, Tang2016, Nakagawa2014, Shao2013}. We assume that the resting microglia become activated to each phenotype at a rate that is influenced by the presence of cytokines. Activation of the microglia to the M1 phenotype is positively influenced by the concentration of pro-inflammatory cytokines \cite{Orihuela2016, Shao2013, Nakagawa2014, Vaughan2018, Wang2012} and negatively influenced by the concentration of anti-inflammatory cytokines \cite{Vaughan2018, Nakagawa2014}. We use the Hill function $H(P)$ to represent the saturating promotion of M1 microglia by the pro-inflammatory cytokines \cite{Byrne2014, Kleiner2013}. Likewise, we use the Hill function $\hat{H}(A)$ to represent the saturating inhibition of M1 microglia by the anti-inflammatory cytokines. Similar terms are used to represent the saturating promotion of M2 microglia by anti-inflammatory cytokines \cite{Orihuela2016, Vaughan2018, Shao2013, Nakagawa2014, Liu2016} and the saturating inhibition of M2 microglia by the pro-inflammatory cytokines \cite{Tang2016}. 

\subsubsection*{Microglia phenotype switching}
There is evidence that microglia may switch phenotypes once activated; however, the switching from the M2 to M1 phenotype has been cited as an area for further research \cite{Boche2013}. The proposed model includes bidirectional switching, so that we may analyze the effects of all possible interactions. To this end, we assume that once activated, microglia may switch from the M1 phenotype to the M2 phenotype \cite{Hu2015, Shao2013, Tanaka2015, Vaughan2018, Nakagawa2014, Cherry2014, Boche2013, Guruswamy2017, Qin2019, Orihuela2016, Zhao2017} and from the M2 phenotype to the M1 phenotype \cite{Orihuela2016, Hu2015, Tanaka2015, Nakagawa2014,  Guruswamy2017, Qin2019, Zhao2017}. Since switching from M1 to M2 is positivity influenced by the anti-inflammatory cytokines \cite{Hu2015,Shao2013}, we multiply this term by the Hill function $H(A)$. Likewise, since switching from M2 to M1 is positively influenced by the pro-inflammatory cytokines \cite{Orihuela2016, Hu2015, Tanaka2015}, we multiply the corresponding term by $H(P)$.


\subsection{Cytokines}
The equations describing the concentration changes of pro-inflammatory ($P$) and anti-inflammatory ($A$) cytokines are as follows:
\begin{equation} \label{eq:P}
\frac{dP}{dt} = \underbrace{k_{P} \cdot M1 \cdot H(P) \cdot \hat{H}(A)}_\text{pro-inflammatory cytokine production} - \underbrace{\mu_{P} \cdot P}_\text{natural decay}
\end{equation}
\begin{equation} \label{eq:A}
\frac{dA}{dt} = \underbrace{k_{A} \cdot M2 \cdot H(A) \cdot \hat{H}(P)}_\text{anti-inflammatory cytokine production} - \underbrace{\mu_{A} \cdot A}_\text{natural decay}
\end{equation}
where $k_P$, $k_A$, $\mu_P$, and $\mu_A$ are constant parameters related to the production and decay of pro-inflammatory and anti-inflammatory cytokines. The Hill functions $H$ and $\hat{H}$ take the same form as in \eqref{eq:H} and \eqref{eq:Hhat}, respectively. The following subsection details the model terms relating to cytokine production, and Table \ref{tab: ParameterTable} lists the relevant parameter descriptions, units, and nominal values.

\subsubsection*{Cytokine production}
We assume that pro-inflammatory cytokines are produced by M1 cells at a rate $k_P$, anti-inflammatory cytokines are produced by M2 cells at a rate $k_A$, and their production is influenced by the presence of both the pro- and anti-inflammatory cytokines in the system \cite{Orihuela2016, Hu2015, Taylor2013, Tang2016, Hao2016, Shao2013, Nakagawa2014}. In particular, the concentration of pro-inflammatory cytokines supports additional pro-inflammatory cytokines production and suppresses the production of anti-inflammatory cytokines \cite{Shao2013, Orihuela2016, Hu2015, Taylor2013, Tang2016}, which we model through the use of the Hill functions $H(P)$ in $\eqref{eq:P}$ and $\hat{H}(P)$ in $\eqref{eq:A}$. Further, the presence of anti-inflammatory cytokines encourages more anti-inflammatory cytokines to be produced and suppresses the production of pro-inflammatory cytokines \cite{Shao2013, Orihuela2016, Hu2015, Taylor2013, Tang2016}, which we model through the terms $H(A)$ in $\eqref{eq:A}$ and $\hat{H}(A)$ in $\eqref{eq:P}$. The Hill functions account for the saturating effects of these interactions \cite{Byrne2014, Kleiner2013}.


\subsection{Model Summary and Simulation Results}
In summary, the proposed model describing the interactions between the M1 and M2 microglia phenotypes and pro-inflammatory and anti-inflammatory cytokines comprises Equations \eqref{eq:M1}, \eqref{eq:M2}, \eqref{eq:P}, and \eqref{eq:A}. The complete system of coupled nonlinear differential equations is given by 
\begin{equation}
\begin{split}
\frac{dM1}{dt} &= a \cdot k_{M1} \cdot H(P) \cdot \hat{H}(A) - s_{M1 \rightarrow M2} \cdot H(A) \cdot M1 + s_{M2 \rightarrow M1} \cdot H(P)  \cdot M2 - \mu_{M1} \cdot M1 \\
\frac{dM2}{dt} &= a \cdot k_{M2} \cdot H(A) \cdot \hat{H}(P) + s_{M1 \rightarrow M2} \cdot H(A) \cdot M1 - s_{M2 \rightarrow M1} \cdot H(P)  \cdot M2 - \mu_{M2} \cdot M2 \\
\frac{dP}{dt} &= k_{P} \cdot M1 \cdot H(P) \cdot \hat{H}(A) - \mu_{P} \cdot P \\
\frac{dA}{dt} &= k_{A} \cdot M2 \cdot H(A) \cdot \hat{H}(P) - \mu_{A} \cdot A
\end{split}
\label{eq:Model}
\end{equation}
with $H$ and $\hat{H}$ defined as in \eqref{eq:H} and \eqref{eq:Hhat}, respectively.

{\renewcommand{\arraystretch}{1.5}
\begin{table}[t!]
\centering
 \begin{tabular}{||c|c| c|c| c||} 
 \hline
 Index & Parameter & Description & Nominal Value & Units \\ [0.5ex] 
 \hline\hline 
  $1$ & $a$ & Number of resting microglia & $1000$ & cells\\ 
   \hline
$2$ & $k_{M1}$ & Activation rate of microglia to $M1$  & $0.44$ & $\frac{1}{\text{hours}}$   \\  
 \hline
$3$ &  $k_{M2}$ & Activation rate of microglia to $M2$  & $0.65$ & $\frac{1}{\text{hours}}$  \\
 \hline
 $4$ & $k_{P}$ & Production rate of $P$ & $0.01$ & $\frac{pg/ml}{\text{hours} \cdot\text{cells}}$ \\ 
 \hline
 $5$ & $k_{A}$ & Production rate of $A$ &  $0.006$& $\frac{pg/ml}{\text{hours} \cdot \text{cells}}$ \\
 \hline
$6$ & $n_{P}$ & Hill coefficient for $P$ in $H(P)$ & $0.5$ & unitless \\
 \hline
$7$ &  $K_{P}$ &  Half-maximal concentration of $P$ & $10$ & $\frac{pg}{ml}$ \\
\hline
$8$ & $n_{A}$ & Hill coefficient for $A$ in $H(A)$  & $0.5$ & unitless \\
 \hline
  $9$ &  $K_{A}$ &  Half-maximal concentration of $A$   & $10$ & $\frac{pg}{ml}$ \\ 
 \hline
 $10$ &   $s_{M1 \rightarrow M2}$ & Rate of $M1 \rightarrow M2$ switch &  $0.2$ & $\frac{1}{\text{hours}}$ \\
 \hline
$11$ &  $s_{M2 \rightarrow M1}$ & Rate of $M2 \rightarrow M1$ switch &  $0.2$ & $\frac{1}{\text{hours}}$  \\
 \hline
 $12$ &  $\mu_{M1}$ & Mortality rate of $M1$ & $0.1$ & $\frac{1}{\text{hours}}$  \\
 \hline
$13$ &    $\mu_{M2}$ & Mortality rate of $M2$ & $0.1$ & $\frac{1}{\text{hours}}$ \\
 \hline
 $14$ &   $\mu_{A}$ & Natural decay rate of $A$ & $0.1$ & $\frac{1}{\text{hours}}$ \\
 \hline
 $15$ &   $\mu_{P}$ & Natural decay rate of $P$ & $0.1$ & $\frac{1}{\text{hours}}$  \\ 
 \hline
 $16$ & $N_{A}$ & Hill coefficient for $A$ in $\hat{H}(A)$ & $0.5$ & unitless \\
 \hline
$17$ & $N_{P}$ & Hill coefficient for $P$ in $\hat{H}(P)$ & $0.5$ & unitless \\
  \hline
\end{tabular}
\caption{Indices, descriptions, nominal values, and units of the constant parameters in Model \eqref{eq:Model}.}
  \label{tab: ParameterTable}
\end{table}

Table \ref{tab: ParameterTable} lists the index, description, nominal value, and unit for each parameter included in Model \eqref{eq:Model}. Nominal parameter values  were chosen to obtain model outputs consistent with experimental findings \cite{Li2018, Li2019, Liu2017, Jiang2018, Li2012, Taj2016, Liu2018, Liu2019}. In particular, simulations with nominal parameter values reflect dominance of M2 microglia early in the inflammatory process, followed by an eventual takeover of the M1 microglia \cite{Hu2012, Tang2016}. 

Numerical simulations were performed using MATLAB\textsuperscript{\textregistered} programming language. Specifically, \texttt{ode15s} was utilized to compute the numerical solution of Model \eqref{eq:Model} using the nominal parameters in Table \ref{tab: ParameterTable} and the initial conditions $M1(0)=100$ cells, $M2(0)=100$ cells, $P(0)=10 \ \frac{pg}{ml}$, and $A(0)=10 \ \frac{pg}{ml}$. Figure~\ref{fig:ModelOutput} shows the resulting model output for the numbers of M1 and M2 microglia and the concentrations of pro-inflammatory and anti-inflammatory cytokines, as well as the ratio of M1 to M2 cells ($M1:M2$) and the total number of activated microglia ($M1 + M2$), over a $72$ hour time period. 

Note in Figure~\ref{fig:ModelOutput} that the M2 phenotype dominates until around $17$ hours. After this time period, the ratio becomes greater than one, indicating that there are more M1 microglia than M2 microglia. At $72$ hours, the ratio of M1 to M2 microglia is approximately $1.34$. The minimum ratio is approximately $0.7875$ and occurs at $2.4$ hours. We also observe that after a short time the concentration of pro-inflammatory cytokines is significantly larger than that of the anti-inflammatory cytokines. We verify that these values are within experimentally observed ranges of cytokines post ischemic stroke \cite{Wang2012, Kleiner2013, Ferrarese1999}. While not shown, note that when the switching from M1 to M2 microglia is turned off, the M1 microglia dominate from the beginning, and the M2 microglia and anti-inflammatory cytokines approach zero. Similarly, when the switching from M2 to M1 is turned off, the M2 microglia and anti-inflammatory cytokines dominate.

\begin{figure}[p] 
  \begin{subfigure}[b]{0.5\linewidth}
    \centering
    \includegraphics[width=1\linewidth]{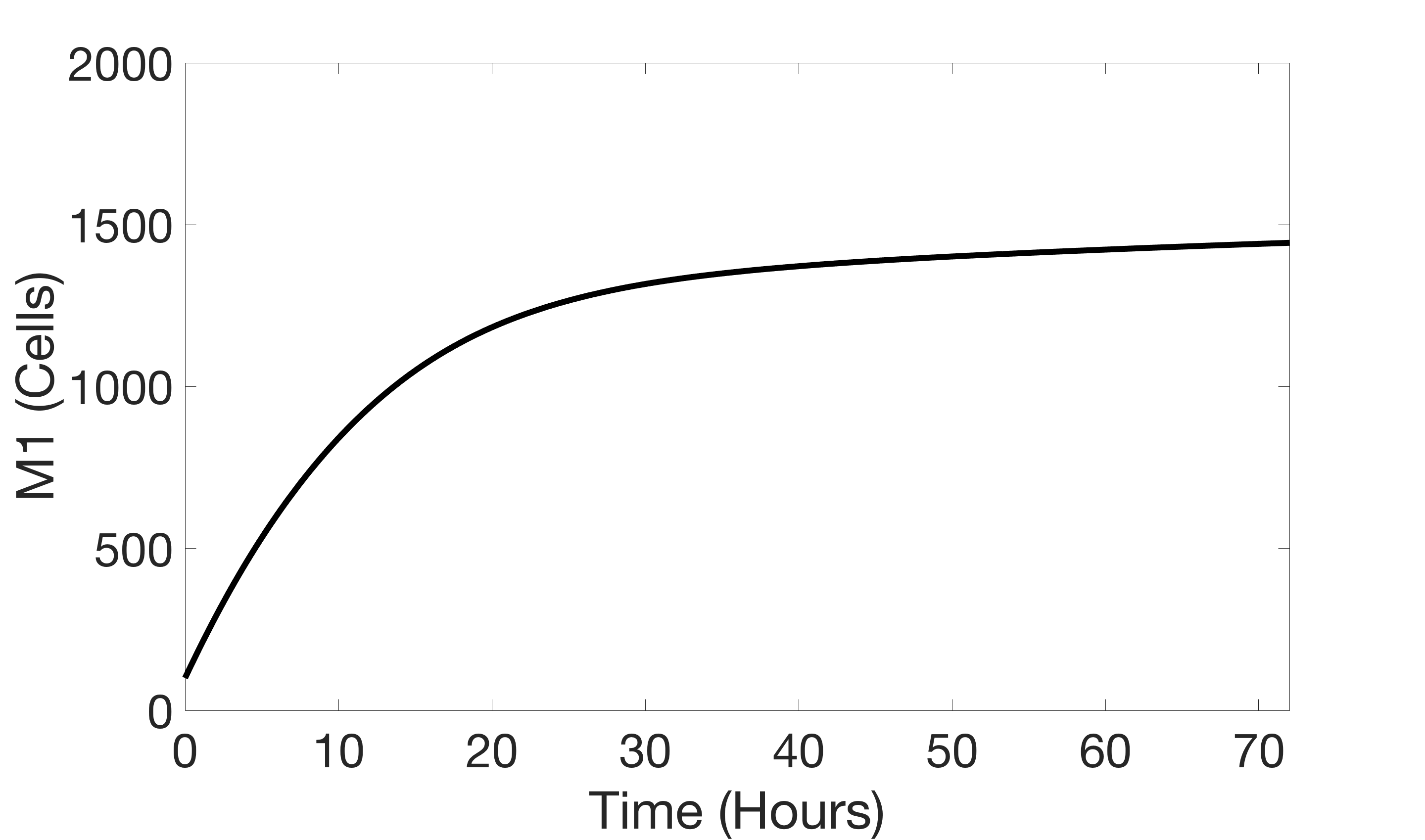} 
    \caption{M1 microglia.} 
        \label{fig: M1output}
    \vspace{4ex}
  \end{subfigure} 
  \begin{subfigure}[b]{0.5\linewidth}
    \centering
    \includegraphics[width=1\linewidth]{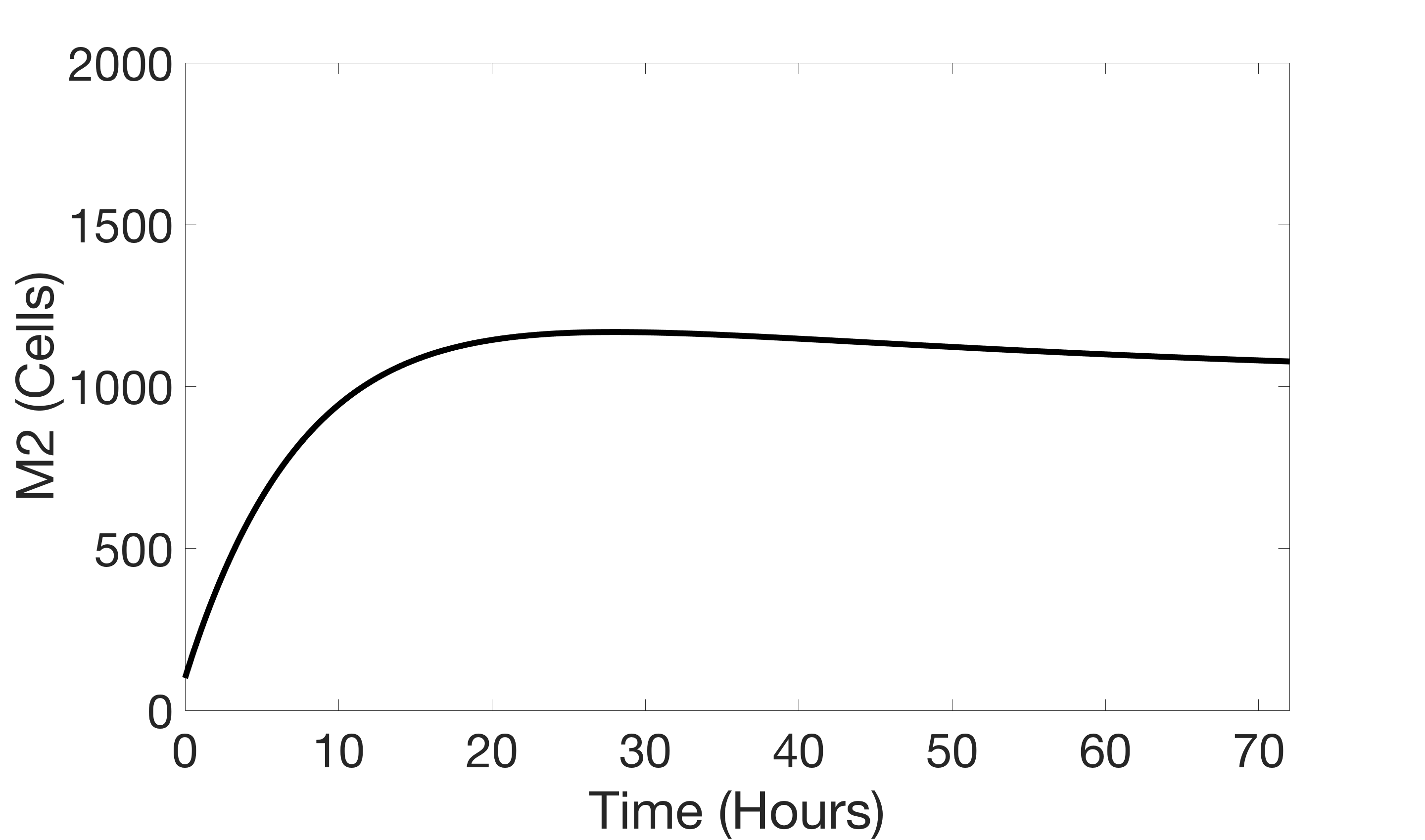} 
    \caption{M2 microglia.} 
            \label{fig: M2output}
    \vspace{4ex}
  \end{subfigure} 
    \begin{subfigure}[b]{0.5\linewidth}
    \centering
    \includegraphics[width=1\linewidth]{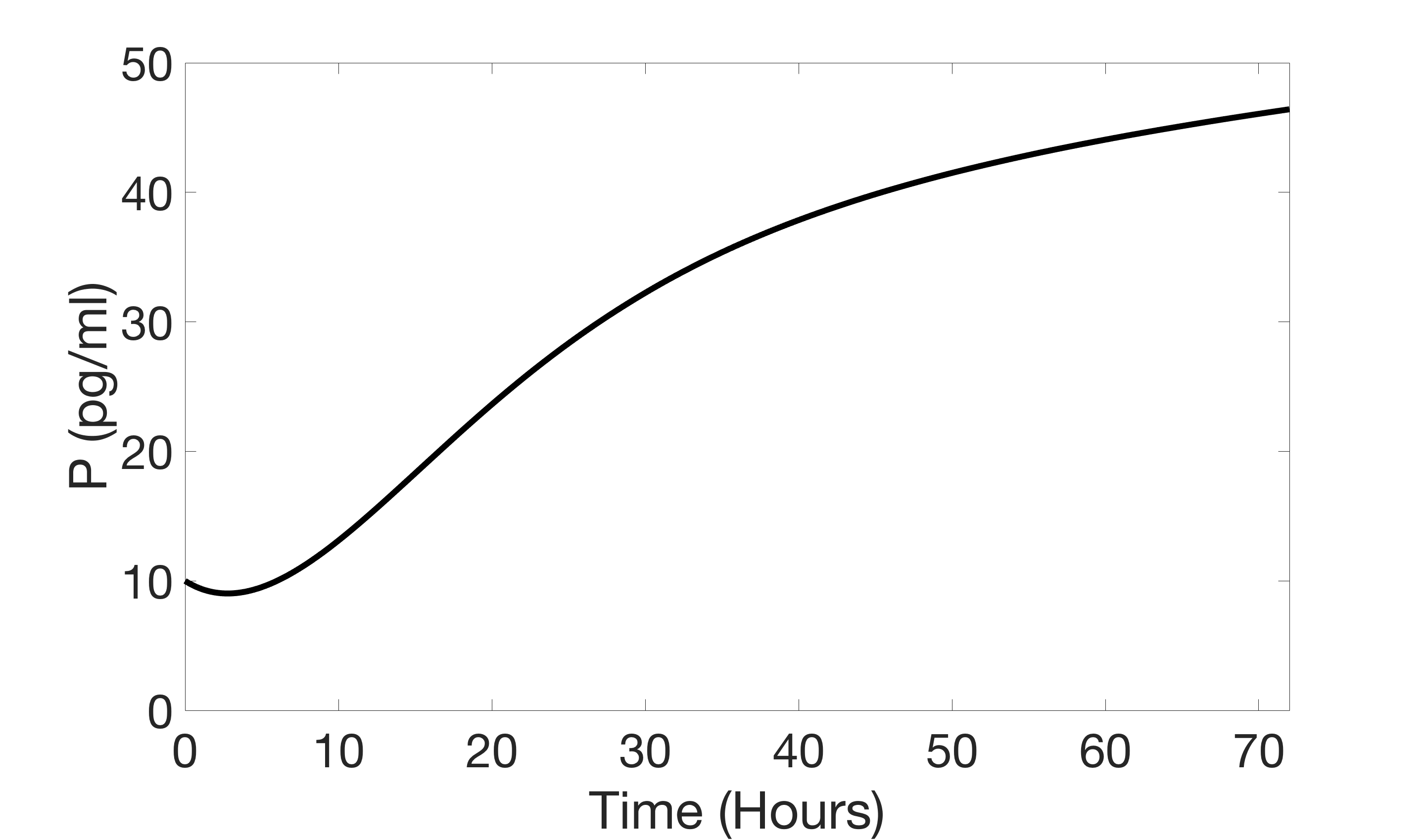} 
    \caption{Pro-inflammatory cytokine concentration.} 
    \vspace{4ex}
  \end{subfigure}
    \begin{subfigure}[b]{0.5\linewidth}
    \centering
    \includegraphics[width=1\linewidth]{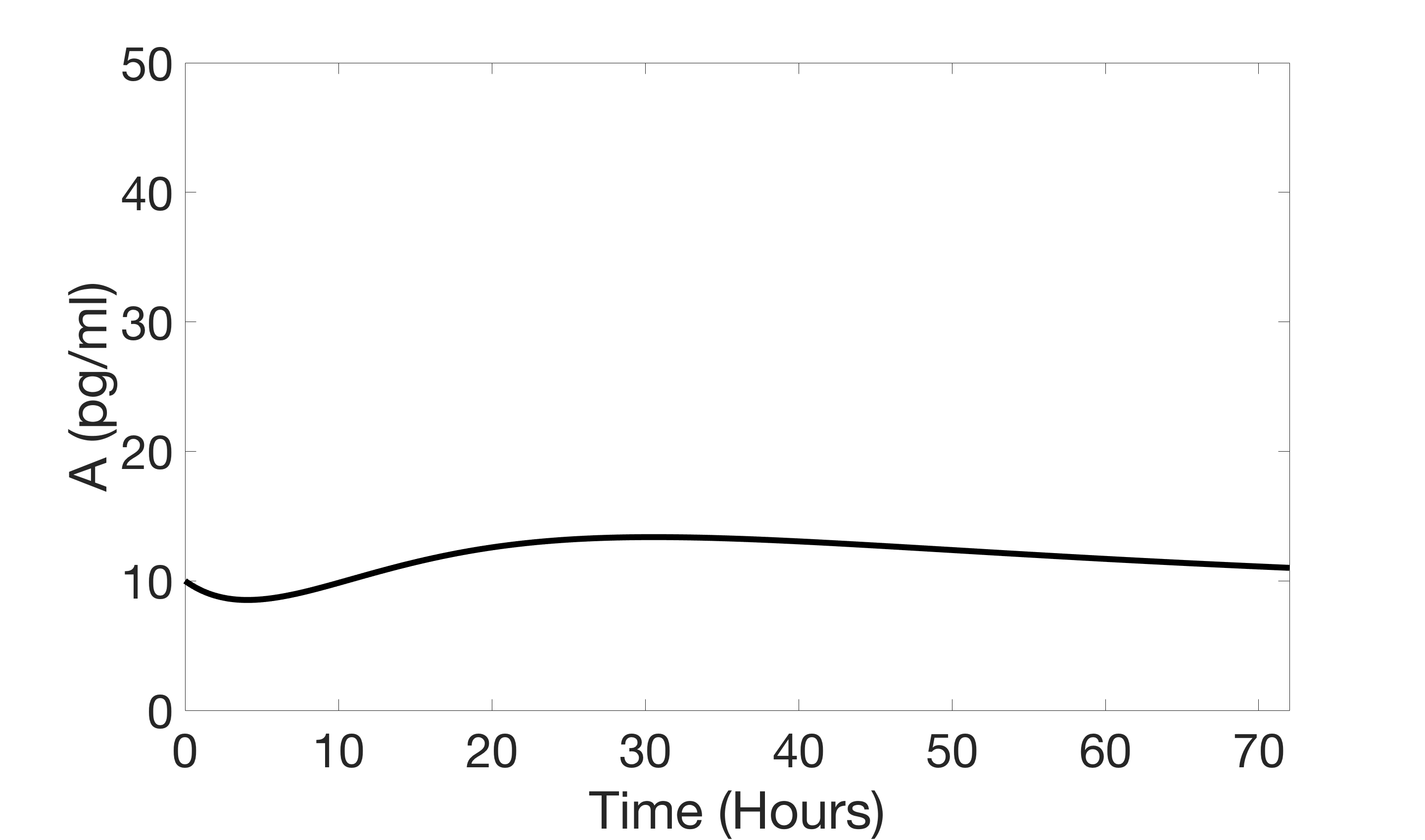} 
    \caption{Anti-inflammatory cytokine concentration.} 
    \vspace{4ex}
  \end{subfigure} 
      \begin{subfigure}[b]{0.5\linewidth}
    \centering
    \includegraphics[width=1\linewidth]{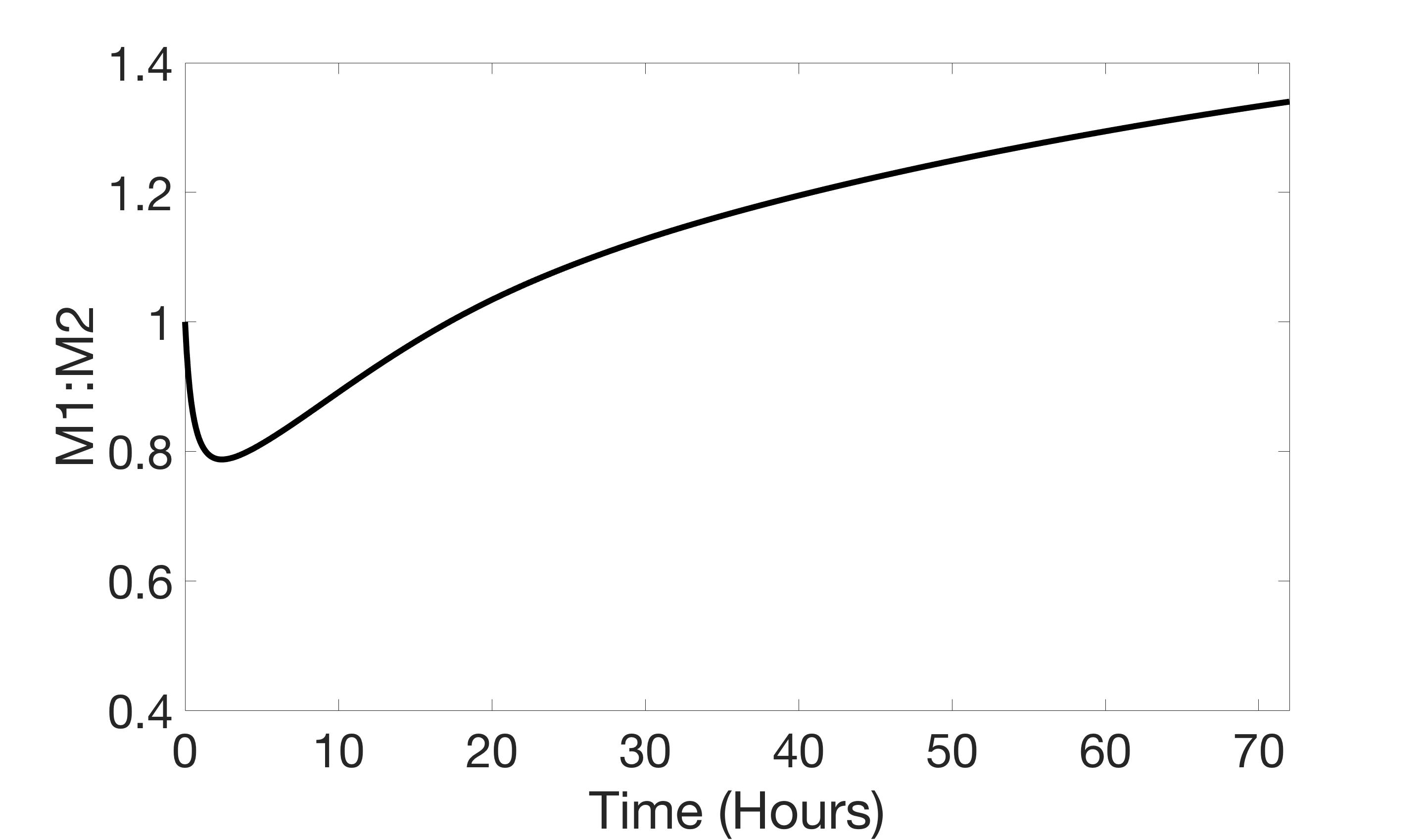} 
    \caption{Ratio of M1 to M2 cells. } 
    \vspace{4ex}
  \end{subfigure} 
      \begin{subfigure}[b]{0.5\linewidth}
    \centering
    \includegraphics[width=1\linewidth]{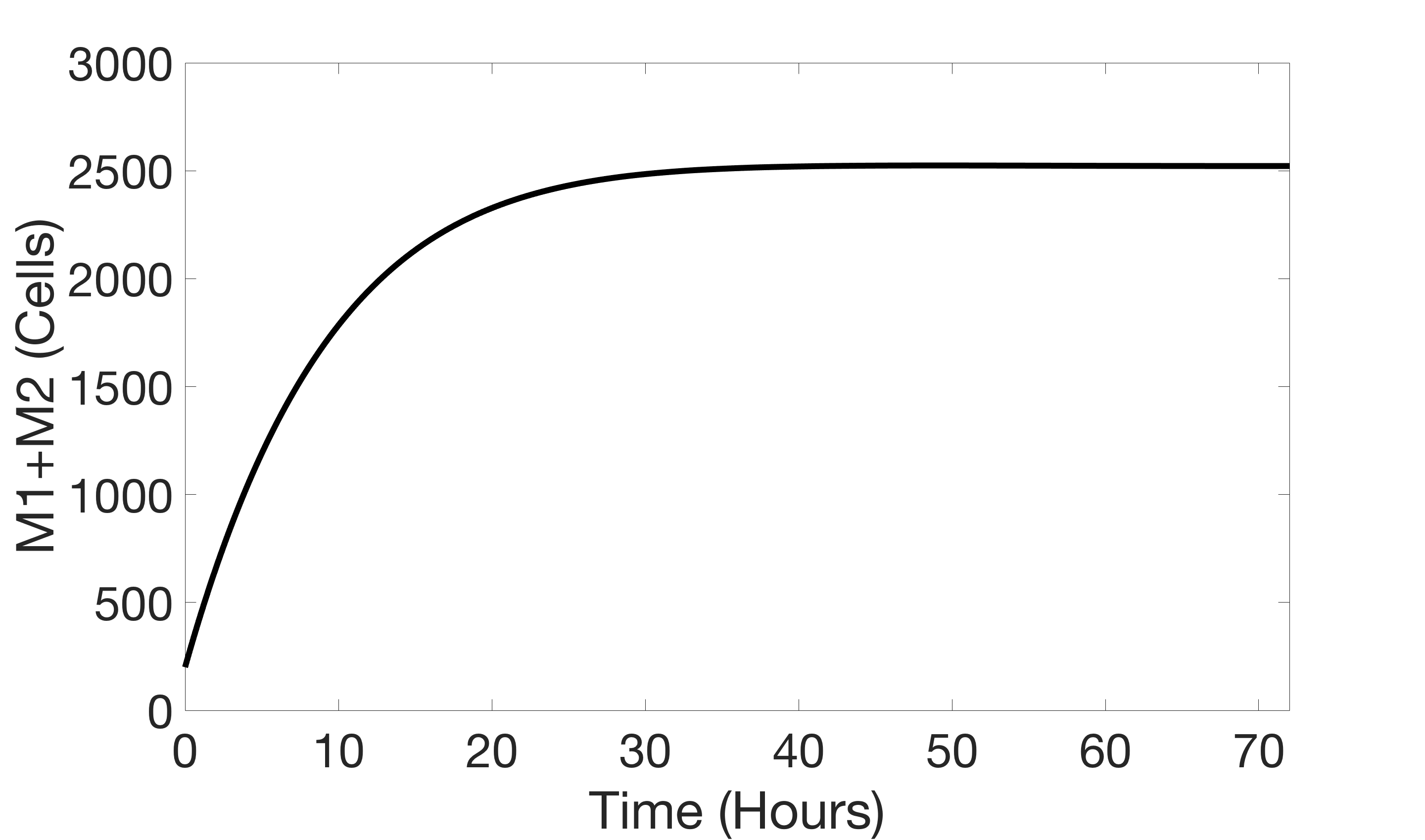} 
    \caption{Total activated microglia.} 
    \vspace{4ex}
  \end{subfigure} 
  \caption{Numerical solution to the model \eqref{eq:Model} over the time interval $[0,72]$ hours using the parameter values in Table \ref{tab: ParameterTable} and initial conditions $M1(0)=100$ cells, $M2(0)=100$ cells, $P(0)=10 \ \frac{pg}{ml}$, and $A(0)=10 \ \frac{pg}{ml}$. The plots in (a)-(d) depict the four model states, while (e) and (f) show the ratio of M1 to M2 microglia and the total number of activated microglia, respectively. }
  \label{fig:ModelOutput}
\end{figure}

\section{Sensitivity Analysis} \label{sec: Sensitivity Analysis}
Since the behavior of Model \eqref{eq:Model} is greatly influenced by the choice of values for the 17 model parameters, we utilize sensitivity analysis techniques in order to study each parameter's contribution to model output.  In particular, we apply two global sensitivity analysis techniques: the Morris method of elementary effects, and the Sobol method. Global sensitivity approaches aim to quantify how uncertainty and variability in model output can be attributed to uncertainties in the inputs. We summarize Morris elementary effects and Sobol sensitivity analysis below, describing specifically the application to this work; for more details on these methods, see \cite{Saltelli2004, Wentworth2016, Smith2013, Olsen2019, Wu2013}.

For each method, consider the nonlinear input-output relation
\begin{equation}\label{eq:response}
y = f(q), \quad q =  [q_1, \dots, q_{17}]
\end{equation}
where $y$ is a scalar response variable and each $q_i$ is a model parameter whose index $i$ ($i = 1, \dots, 17$) corresponds to the index listed in Table~\ref{tab: ParameterTable}. Since we are interested in how the parameters affect the number of activated microglia, in particular the ratio of M1 to M2 microglia and the total activated microglia, we consider the following two response variables:
\begin{equation} \label{eq:ratio}
f(q) = \int_0^{72} M1(t;q)./M2(t;q) dt
\end{equation}
\begin{equation} \label{eq:sum}
f(q) = \int_0^{72} (M1(t;q) + M2(t;q)) dt
\end{equation}
Parameters are admitted to vary over a specified space. In this study, each parameter's admissible space is taken to be the interval of $80-120\%$ around the nominal value given in Table~\ref{tab: ParameterTable}.
\par


\subsection{Morris Elementary Effects}
The Morris method of elementary effects quantifies the effect of varying one parameter at a time on a model output. The interval $[0, 1]$ is divided into $\ell$ levels, and parameters are randomly sampled from these levels. In this work, we let $\ell = 100$. Parameters are perturbed one at a time by the increment 
\begin{equation}
\Delta = \frac{\ell}{2{(\ell-1)}}. 
\end{equation}
For each parameter $q_i$, we generate two vectors which differ only in the $i$th entry. Parameters are then rescaled to their admissible parameter space, and elementary effects are computed by
\begin{equation}
d_i(q) = \frac{f(q + \Delta e_i) - f(q)}{\Delta}
\end{equation}
where $q$ is the parameter vector and $e_i$ is the $i$th unit vector. This process is repeated for $r=200$ samples, and the absolute mean $\mu_i^*$ and variance $\sigma_i^2$ are computed via the formulas
\begin{equation} \label{eq:mustar}
\mu_i^* = \frac{1}{r} \sum_{j=1}^r |d_i^j| 
\end{equation}
and
\begin{equation} \label{eq:variance}
\sigma_i^2 = \frac{1}{r-1}  \sum_{j=1}^r (d_i^j - \mu_i)^2, \qquad \mu_i = \frac{1}{r} \sum_{j=1}^r d_i^j
\end{equation}
The absolute mean $\mu_i^*$ in \eqref{eq:mustar} provides an estimate of the absolute value of the average of elementary effects over all samples. The variance $\sigma_i^2$ in \eqref{eq:variance} measures how far each elementary effect is from the mean. Since large variances indicate dependence on neighboring input values, the variance gives an estimate of the combined effects of the interactions of each parameter with other parameters. In this work, we use the absolute mean $\mu_i^*$ to rank the sensitivity of the parameters.


\subsection{Sobol Method}

Sobol sensitivity analysis is a variance-based method which quantifies how much of the variability in the model output can be attributed to each individual parameter or parameter interactions. To implement the Sobol method in this work, parameters are randomly sampled from their respective parameter space a total of $50,000$ times. Half of these samples form the rows of a matrix $A$, which has dimensions $25,000 \times 17$ (where 17 is the number of parameters), and the other half form the rows of a matrix $B$, a nonidentical $25,000 \times 17$ matrix. Seventeen additional $25,000 \times 17$ matrices, denoted as $C_1, \dots, C_{17}$, are generated such that each $C_i$ corresponds to the parameter $q_i$ and has its $i$th row taken as the $i$th row of $A$ and its remaining $16$ rows taken from $B$. 

A scalar model output is generated for each row of the matrices $A$, $B$, and $C_i$ for all $i$ by first running a forward simulation of the model with parameter values set to the entries in the respective row and then computing the response variable \eqref{eq:response}. This results in scalar response vectors of size $1\times 25,000$ for each matrix, denoted by $y_A$, $y_B$, and $y_{C_i}$, where
\begin{equation}\label{scalar}
y_A = f(A), \quad y_B = f(B), \quad y_{C_i} = f(C_i)
\end{equation}
and the output function $f$ is taken to be either \eqref{eq:ratio} or \eqref{eq:sum}, respectively.

The first-order sensitivity indices, $S_i$, and total-order sensitivity indices, $S_{T_i}$, are computed using the formulas
\begin{equation} \label{eq:firstorder}
S_i = \frac{\text{var}[\mathbb{E}(Y|q_i)]}{\text{var}(Y)} = \frac{(\frac{1}{M} \cdot y_A \cdot y_{C_i}^T) - f_0^2}{(\frac{1}{M} \cdot y_A \cdot y_A^T) - f_0^2}
\end{equation}
and
\begin{equation} \label{eq:total}
S_{T_i} = 1 - \frac{\text{var}[\mathbb{E}(Y|q_{\sim i})]}{\text{var}(Y)} = 1-\frac{(\frac{1}{M} \cdot y_B \cdot y_{C_i}^T) - f_0^2}{(\frac{1}{M} \cdot y_A \cdot y_A^T) - f_0^2}
\end{equation}
respectively, where
\begin{equation} \label{eq:f0}
f_0^2 = \frac{1}{M} \sum_{j=1}^{M} y^j_A \cdot \frac{1}{M} \sum_{j=1}^{M} y^j_B
\end{equation}
and here $M = 25,000$. The first-order sensitivity indices in \eqref{eq:firstorder} measure the fractional contribution of a single parameter to the output variance, while the total-order sensitivity indices in \eqref{eq:total} measure the contribution of a single parameter and this parameter's interactions with the other parameters to the output variance. We use the total-oder sensitivity indices $S_{T_i}$ to achieve an overall sensitivity ranking of the parameters. 


\subsection{Parameter Sensitivity Rankings}

Figure~\ref{fig: Sensitivity} shows the resulting parameter sensitivity rankings using both the Morris and Sobol methods for the scalar responses given in \eqref{eq:ratio} and \eqref{eq:sum}. Parameters are ranked according to their Morris absolute means, $\mu_i^*$ in \eqref{eq:mustar}, and total-order Sobol sensitivity indices, $S_{T_i}$ in \eqref{eq:total}. Note that the sensitivity rankings are consistent between methods but depend on the response considered. 

As shown in Figure~\ref{fig: ratiosens}, when considering the ratio of M1 to M2 microglia in \eqref{eq:ratio}, the microglia phenotype switching rates $s_{M2 \rightarrow M1}$ and $s_{M1 \rightarrow M2}$ are the most sensitive parameters with respect to both Morris and Sobol sensitivity measures. These are followed by the half-maximal concentrations $K_A$ and $K_P$ of the anti-inflammatory and pro-inflammatory cytokines, respectively. When instead considering the total activated microglia in \eqref{eq:sum}, Figure~\ref{fig: M1M2sens} shows that most of the parameters have a total-order sensitivity index and an absolute mean very close to $0$. The most sensitive parameter for this response is the number of resting microglia $a$, followed by the microglia activation rates $k_{M1}$ and $k_{M2}$ and the microglia mortality rates $\mu_{M1}$ and $\mu_{M2}$.

\begin{figure}[t!] 
  \begin{subfigure}[b]{0.5\linewidth}
    \centering
    \includegraphics[width=1\linewidth]{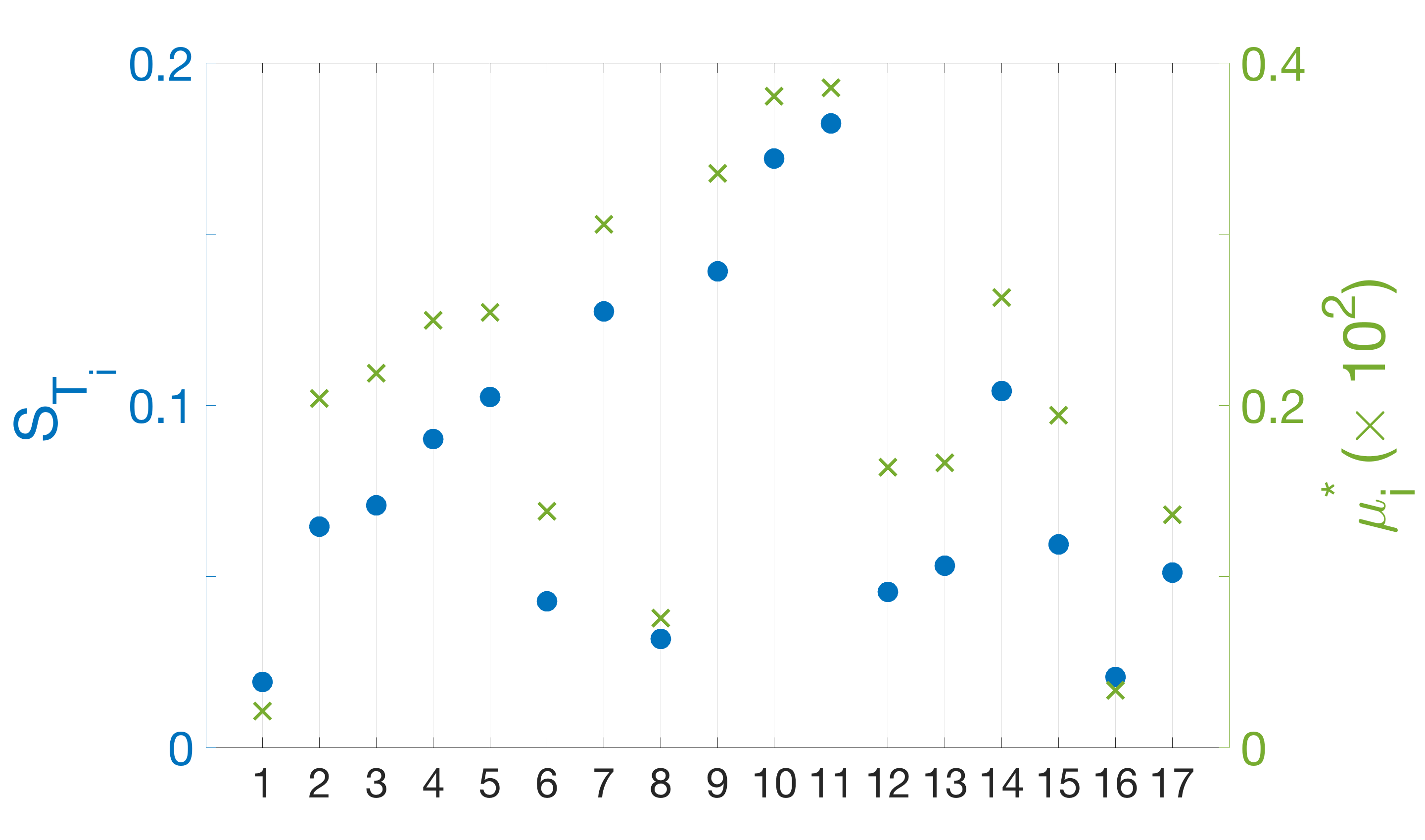} 
    \caption{$f(q) = \int_0^{72} M1(t;q)./M2(t;q) dt$} 
        \label{fig: ratiosens}
         \vspace{4ex}
  \end{subfigure} 
  \begin{subfigure}[b]{0.5\linewidth}
    \centering
    \includegraphics[width=1\linewidth]{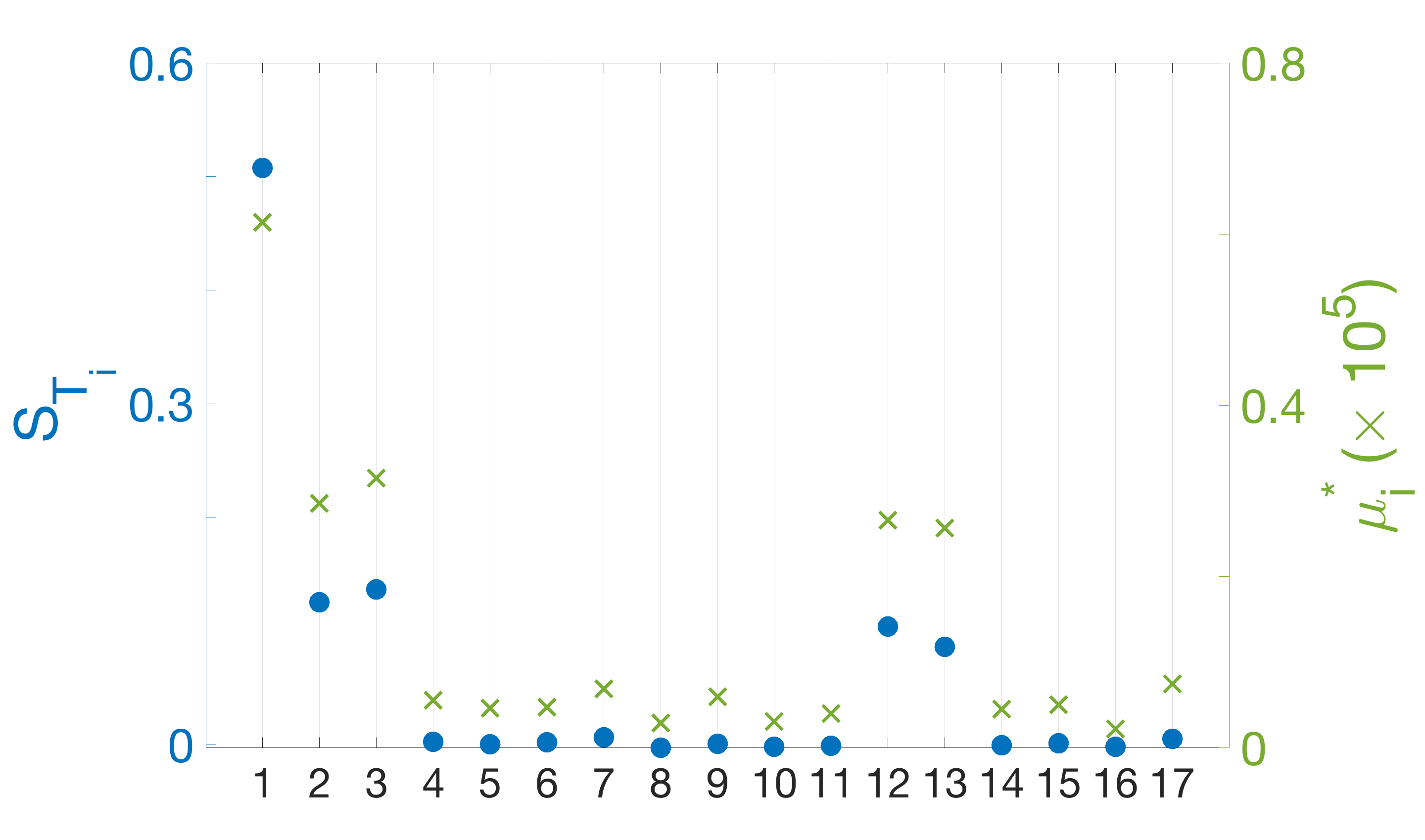} 
    \caption{$f(q) = \int_0^{72} (M1(t;q) + M2(t;q)) dt$} 
    \label{fig: M1M2sens}
     \vspace{4ex}
  \end{subfigure}
    \caption{Total-order sensitivity index $S_{T_i}$ (blue dot) and absolute mean $\mu_i^*$ (green x) for each parameter $q_i$ in Model \eqref{eq:Model} with respect to the scalar response variable $f(q)$. Parameters are labeled using the indices $i$ ($i = 1, \dots, 17$) listed in Table~\ref{tab: ParameterTable}. }
  \label{fig: Sensitivity}
\end{figure}

\section{Modified Model for Neuroprotectant-Based Strategies} \label{sec: Modified Model}

Promoting the activation of M2 microglia while simultaneously suppressing M1 microglia activation has been cited as a possible neuroprotectant strategy to aid during ischemic stroke \cite{Lee2014, Guruswamy2017, Ginsberg2008, Zhao2017, Yenari2010} and has been the subject of several recent experimental studies \cite{Li2018, Li2019, Liu2017, Jiang2018, Li2012, Taj2016, Liu2018, Liu2019}. To simulate the effects of such neuroprotectant-based strategies on microglia activation, we modify Model \eqref{eq:Model} to include time-varying terms to inhibit the production of M1 microglia and bolster the production of M2 microglia cells. 

In introducing these terms, we focus on analyzing the ratio of M1 to M2 microglia over a 72-hour window post stroke, which has been cited as an important time frame for treatment \cite{Lee2014, Guruswamy2017, Hu2015, Wang2007, Taj2016, Zhao2017, Orihuela2016, Dyker1998}. Once being administered at a specified time during this window, we assume that the neuroprotectant will have a saturating effect on the activation of M1 and M2 microglial cells for 24 hours, after which the effects will diminish and activation will return to pre-treatment levels. We further assume that the neuroprotectant should be administered within 15 hours post stroke in order to extend the early dominance of the M2 phenotype over M1 seen in Model \eqref{eq:Model} for as long as possible before the number of M1 cells again dominates.  

Based on these assumptions, we include the following two time-dependent terms:
\begin{equation} \label{eq:N1}
  N_1(t) =
  \begin{cases}
         b + \frac{1-b}{1+e^{\tau_1(t - (t_0 + 5))}} & \text{if $t \leq 24 + t_0$} \\
         b + \frac{1-b}{1+e^{-\tau_1(t - (t_0 + 5)-38)}} & \text{if $t > 24 + t_0$} \\
  \end{cases}
\end{equation}
which acts to inhibit M1 activation, and 
\begin{equation} \label{eq:N2}
  N_2(t) =
  \begin{cases}
         1 + \frac{L}{1+e^{-\tau_2(t - (t_0 + 5))}} & \text{if $t \leq 24 + t_0$} \\
         1 + \frac{L}{1+e^{\tau_2(t - (t_0 + 5)-38)}} & \text{if $t > 24 + t_0$} \\
  \end{cases}
\end{equation}
which acts to promote M2 activation. Each term is a continuous piecewise sigmoidal function, where $L$, $b$, $\tau_1$, and $\tau_2$ are constant parameters which control the shape of the respective sigmoid graphs, and $t_0$ is the onset time of simulated neuroprotectant-based treatment. Note that we account for a time delay of 5 hours in each sigmoid curve reaching its respective point of steepest decline or incline, and we shift the curves when $t > 24 + t_0$ by 38 hours in order to maintain continuity. These terms enter Model \eqref{eq:Model} as multiplicative factors, with $N_1(t)$ multiplying the M1 activation term in \eqref{eq:M1} and $N_2(t)$ multiplying the M2 activation term in \eqref{eq:M2}. Corresponding parameter descriptions and nominal values are given in Table \ref{tab: NeuroTable}.

\begin{table}
\centering
 \begin{tabular}{||c|c| c| c|c||} 
 \hline
Index & Parameter & Description & Nominal Value & Units \\ [0.5ex] 
 \hline\hline
$18$ & $b$ & Minimum value of $N_1$ & $0.3$ & unitless  \\
\hline
$19$ & $\tau_1$ & Steepness of $N_1$ & $1$ & unitless  \\
\hline
$20$ & $\tau_2$ & Steepness of $N_2$ & $1$ & unitless  \\
\hline
$21$ & $L$ & Maximum value of $N_2$ & $0.8$ & unitless  \\
\hline
$22$ & $t_0$ & Time at which treatment is applied & $0-15$ & hours\\
\hline
\end{tabular}
\caption{Indices, descriptions, nominal values, and units of the constant parameters in the time-varying neuroprotectant terms $N_1(t)$ and $N_2(t)$ given in \eqref{eq:N1} and \eqref{eq:N2}, respectively.}
  \label{tab: NeuroTable}
\end{table}

To inhibit the production of M1 microglial cells, $N_1(t)$ has the form of a decreasing sigmoid curve until $24$ hours after the onset of treatment, where we assume that the effects of the neuroprotectant start to wear off. After $24$ hours post-treatment, $N_1(t)$ becomes an increasing sigmoid curve until the M1 activation returns to pre-treatment level. The constant parameter $b$ is the minimum value of $N_1(t)$ and represents how effective $N_1(t)$ is in suppressing the activation of M1 cells. Note that if $b=0$, $N_1(t)$ would completely turn off the activation of resting microglia to the M1 phenotype. Figure~\ref{fig:N1_term} shows $N_1(t)$ for onset times $t_0 = 0$, 5, 10, and 15 hours.

Conversely, to bolster the production of M2 microglia cells, $N_2(t)$ has the form of an increasing sigmoid curve until $24$ hours after the onset of treatment, at which point we assume that the effects start to wear off. After $24$ hours post-treatment, $N_2(t)$ transitions to a decreasing sigmoid curve until the M2 activation returns to pre-treatment level. The constant parameter $L$ is the maximum value of $N_2(t)$ and represents how effective $N_2(t)$ is in bolstering the activation of M2 cells. Note that if $L=1$, $N_2(t)$ would double the activation of resting microglia to the M2 phenotype. Figure~\ref{fig:N2_term} plots $N_2(t)$ when $t_0 = 0$, 5, 10, and 15 hours.


\subsection{Modified Model Summary and Simulation Results}

In summary, the modified model for suppressing the activation of the M1 phenotype and bolstering the activation of M2 microglia phenotype is given by the following system of equations: 
\begin{equation}
\begin{split}
\frac{dM1}{dt} &= N_1(t) \cdot a \cdot k_{M1} \cdot H(P) \cdot \hat{H}(A) - s_{M1 \rightarrow M2} \cdot H(A) \cdot M1 + s_{M2 \rightarrow M1} \cdot H(P)  \cdot M2 - \mu_{M1} \cdot M1 \\
\frac{dM2}{dt} &= N_2(t) \cdot a \cdot k_{M2} \cdot H(A) \cdot \hat{H}(P) + s_{M1 \rightarrow M2} \cdot H(A) \cdot M1 - s_{M2 \rightarrow M1} \cdot H(P)  \cdot M2 - \mu_{M2} \cdot M2 \\
\frac{dP}{dt} &= k_{P} \cdot M1 \cdot H(P) \cdot \hat{H}(A) - \mu_{P} \cdot P \\
\frac{dA}{dt} &= k_{A} \cdot M2 \cdot H(A) \cdot \hat{H}(P) - \mu_{A} \cdot A
\end{split}
\label{eq:ModifiedModel}
\end{equation}

Figure~\ref{fig:NeuroModel015} plots the neuroprotectant terms $N_1(t)$ and $N_2(t)$ over $72$ hours when the simulated treatment is administered at $t_0 = 0$, 5, 10, and 15 hours, along with the corresponding plots of M1 and M2 cells, the ratio of M1 to M2 microglia, and total activated microglia resulting from Model \eqref{eq:ModifiedModel}. Note that in the presence of the neuroprotectant terms, the ratio of M1 to M2 microglia remains under one until around $60$ hours regardless of treatment onset time. Applying the neuroprotectant terms right away (i.e., $t_0 = 0$) results in the lowest ratio of M1 to M2 microglia. This minimum occurs at $11.5$ hours and is about $0.5$, indicating that, at this time, there are around twice as many M2 microglial cells as there are M1. After this minimum is achieved, the ratio of microglial cells increases and ends at $72$ hours with the highest ratio of all the onset times considered. For the other onset times shown, a slightly different behavior is observed: The ratios increase prior to treatment onset and then decrease less drastically and level off. Ratios then increase and, by the end of the 72 hour period, all end up around one on an upward trend.

While not shown here, we note that onset times past the 15 hour mark (i.e., $t_0 > 15$) yield a similar behavior seen in Figure~\ref{fig:Neuro_ratio} when $t_0 = 5$, 10, and 15 hours; however, the ratio of M1 to M2 microglial cells no longer stays under one prior to the simulated treatment onset. Instead, the ratios when $t_0 > 15$ increase prior to the onset time, reach a maximum ratio greater than one, and then decrease and level off. When the effects of the neuroprotectant begin to wear off, the ratios increase and end on an upward trend.

\begin{figure}[p] 
        \begin{subfigure}[b]{1\linewidth}
    \centering
    \includegraphics[width=0.6\linewidth]{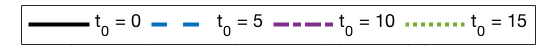} 
    \vspace{4ex}
  \end{subfigure} 
  \begin{subfigure}[b]{0.5\linewidth}
    \centering
    \includegraphics[width=1\linewidth]{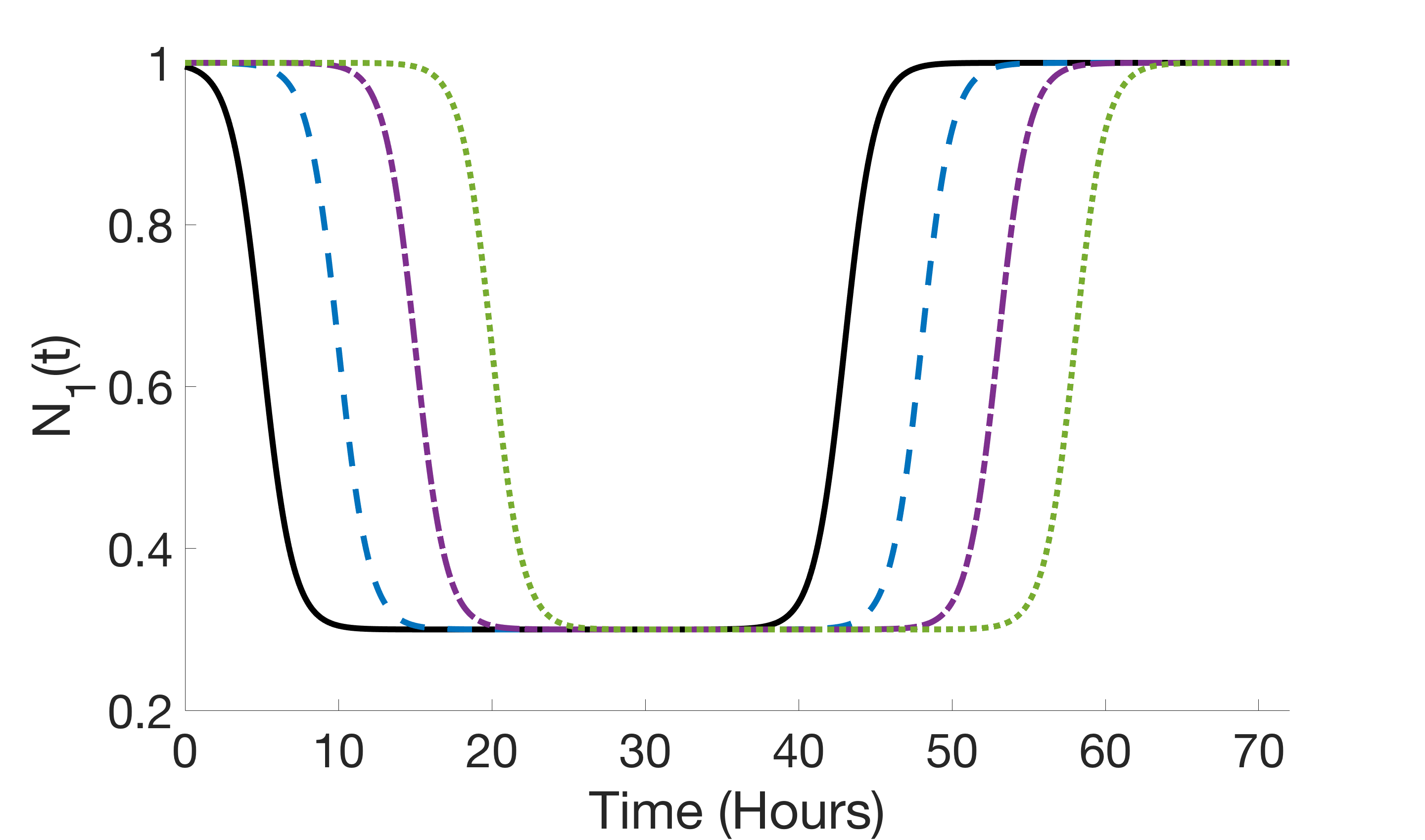} 
    \caption{M1 activation inhibitor.} 
    \label{fig:N1_term}
    \vspace{4ex}
  \end{subfigure} 
  \begin{subfigure}[b]{0.5\linewidth}
    \centering
    \includegraphics[width=1\linewidth]{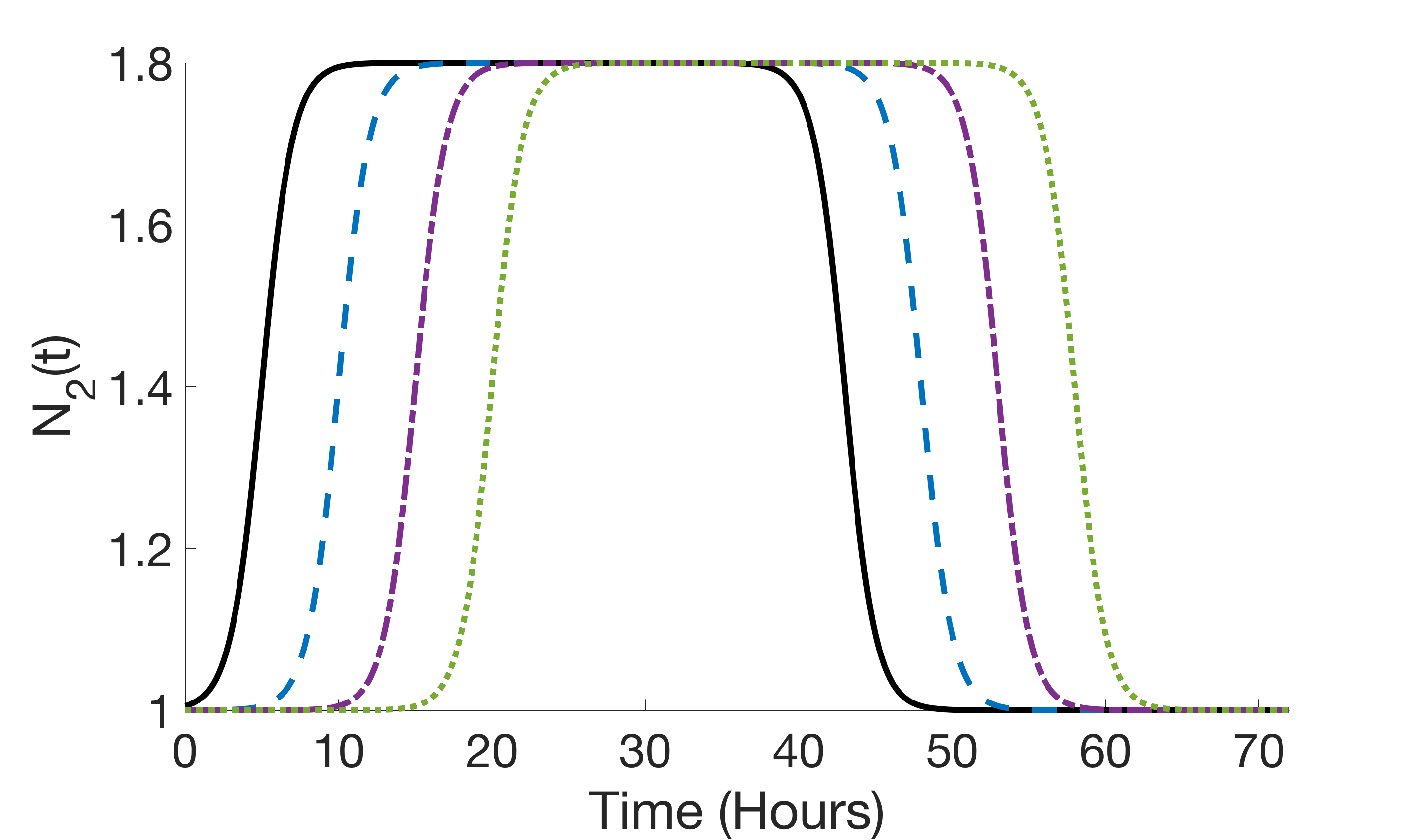} 
    \caption{M2 activation promoter.} 
    \label{fig:N2_term}
    \vspace{4ex}
  \end{subfigure} 
    \begin{subfigure}[b]{0.5\linewidth}
    \centering
    \includegraphics[width=1\linewidth]{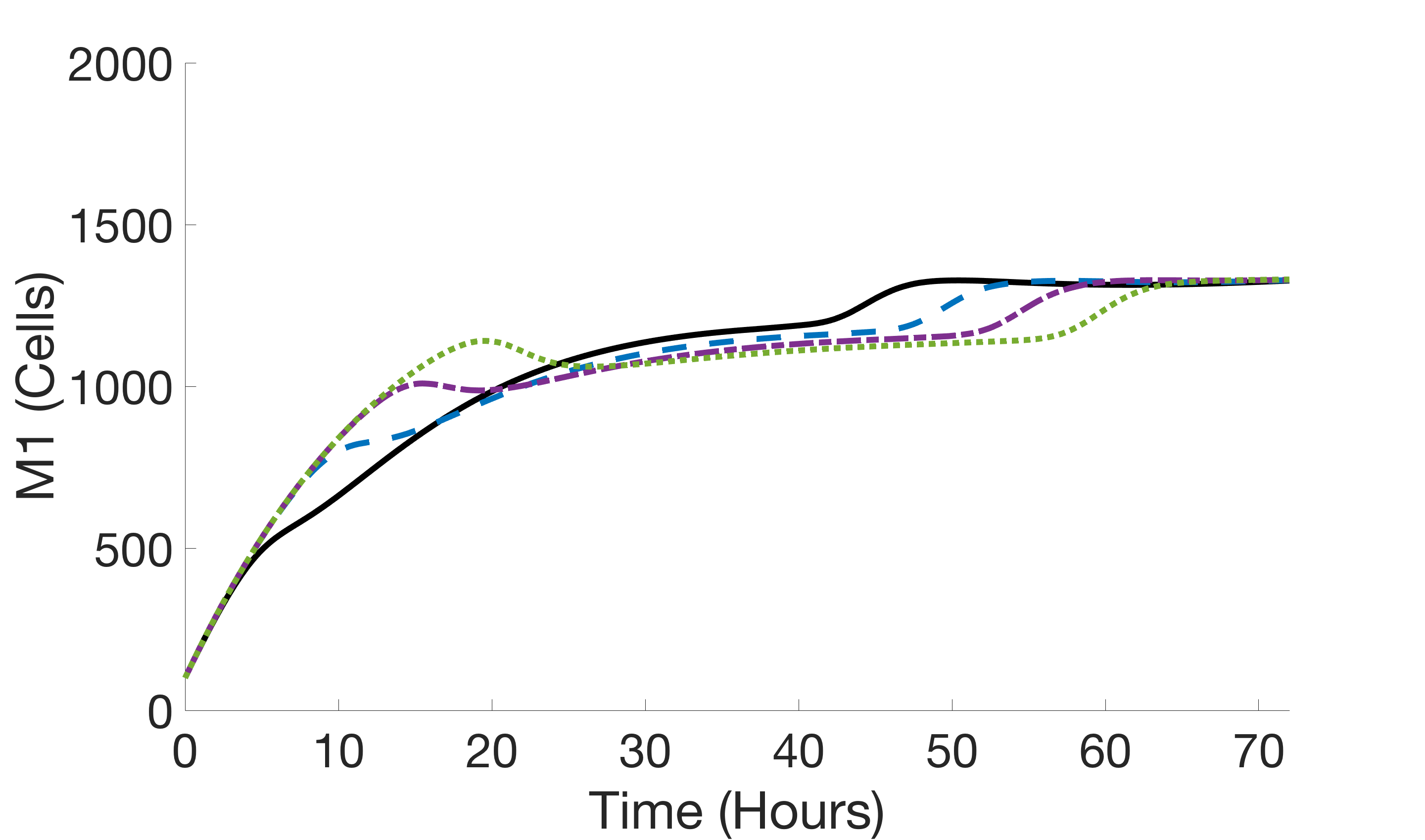} 
    \caption{M1 microglia.} 
    \vspace{4ex}
  \end{subfigure}
    \begin{subfigure}[b]{0.5\linewidth}
    \centering
    \includegraphics[width=1\linewidth]{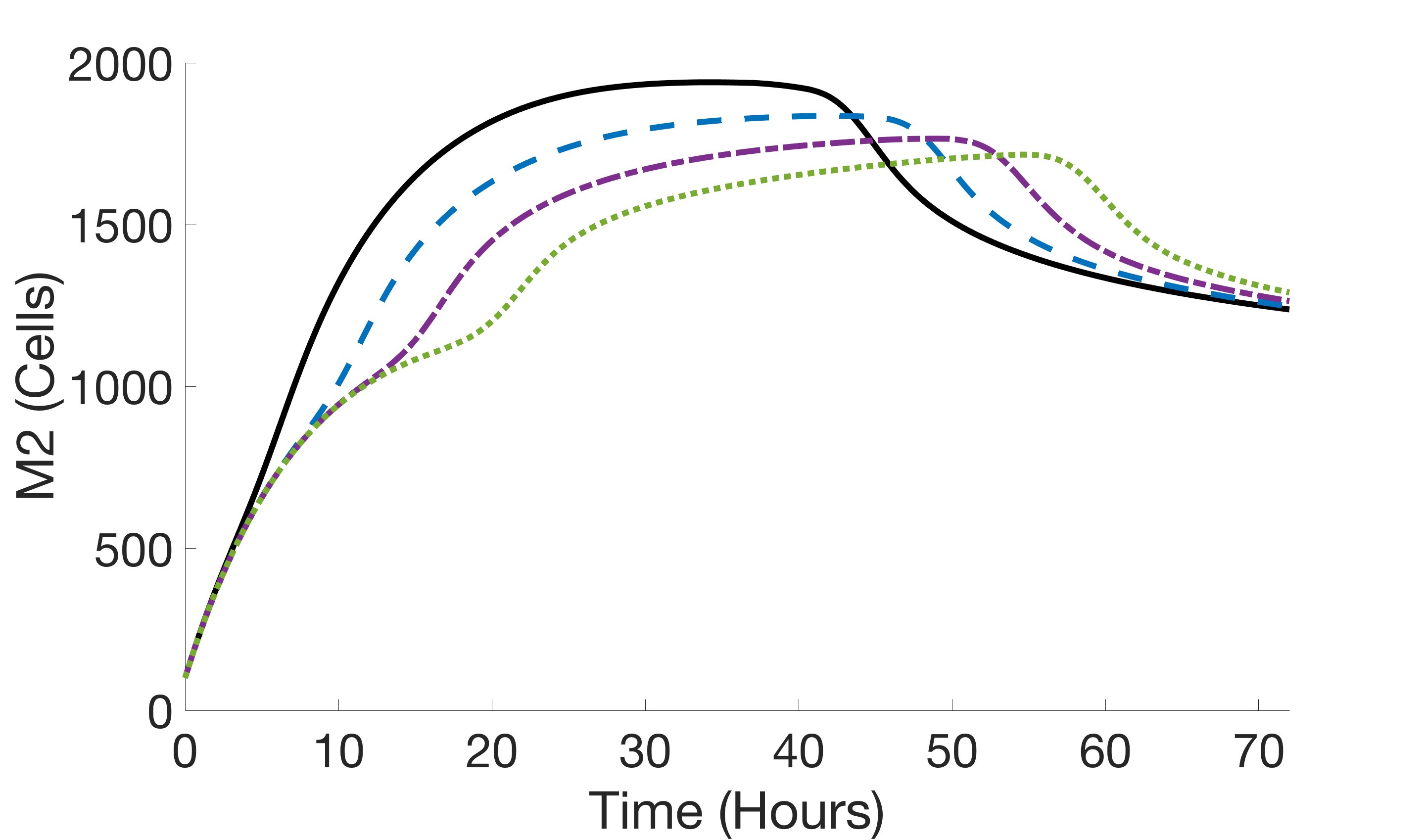} 
    \caption{M2 microglia.} 
    \vspace{4ex}
  \end{subfigure} 
      \begin{subfigure}[b]{0.5\linewidth}
    \centering
    \includegraphics[width=1\linewidth]{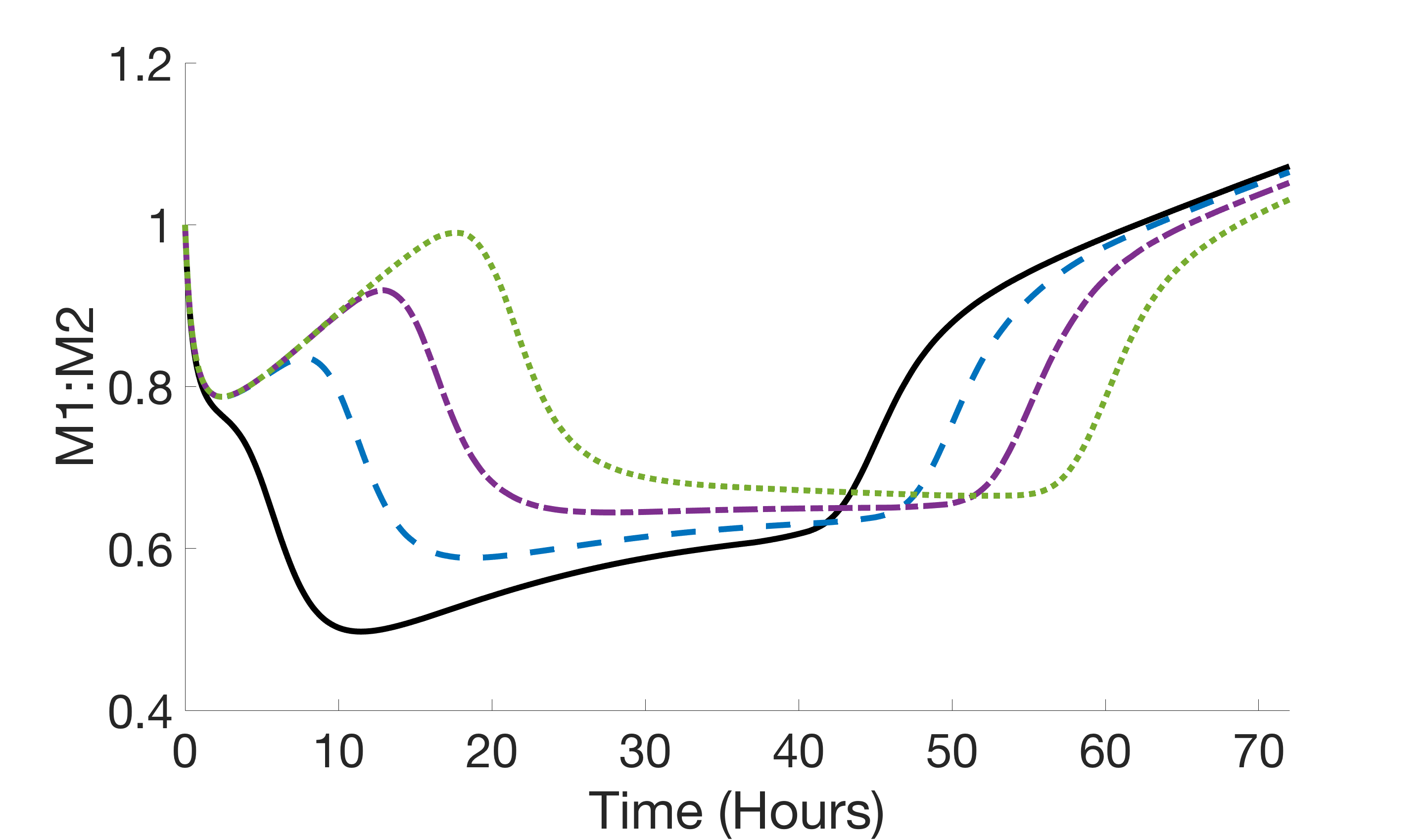} 
    \caption{Ratio of M1 to M2 cells. } 
    \label{fig:Neuro_ratio}
    \vspace{4ex}
  \end{subfigure} 
      \begin{subfigure}[b]{0.5\linewidth}
    \centering
    \includegraphics[width=1\linewidth]{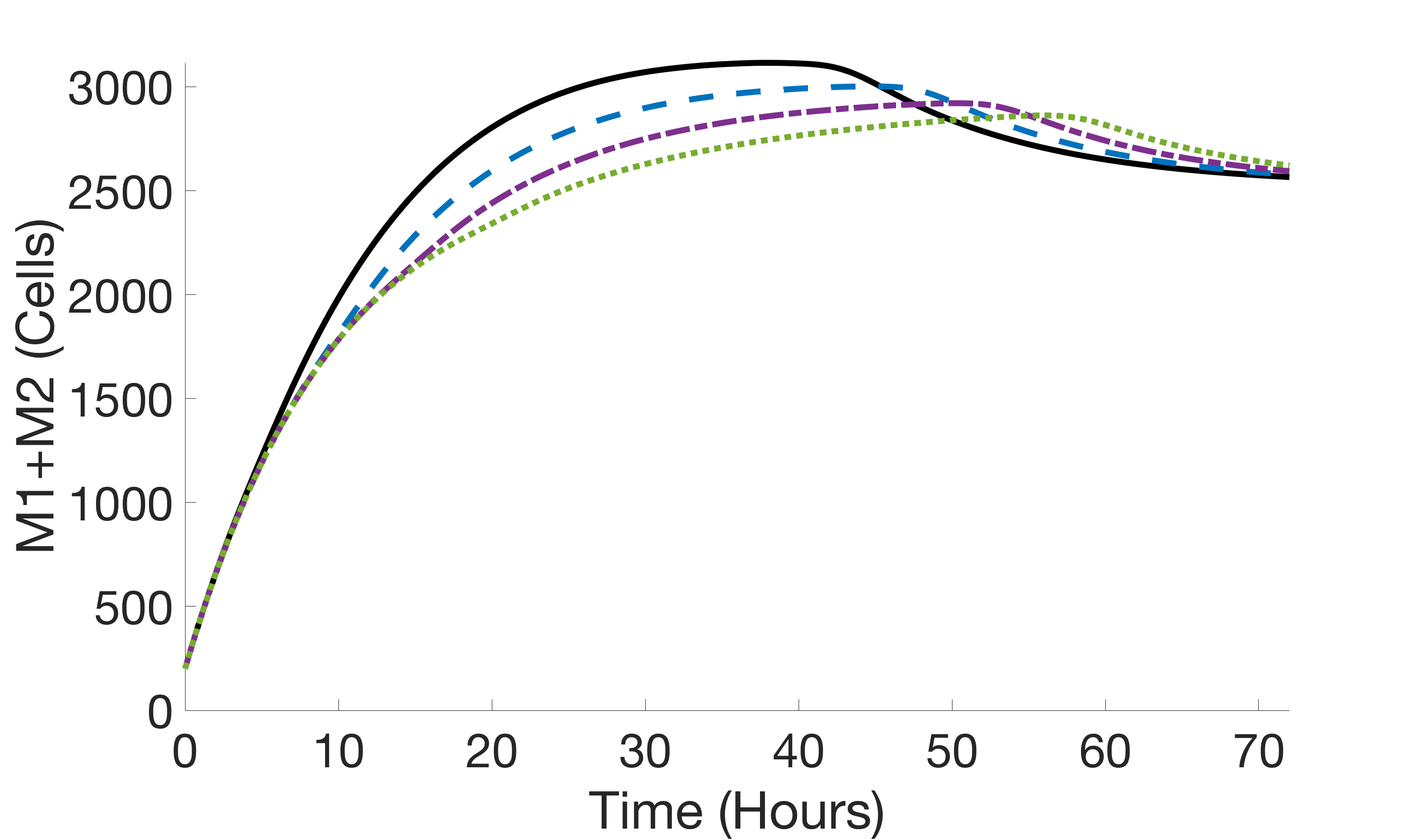} 
    \caption{Total activated microglia.} 
    \vspace{4ex}
  \end{subfigure} 
  \caption{Numerical solutions to the modified model in \eqref{eq:ModifiedModel} over $[0,72]$ hours using the nominal parameters values given in Tables \ref{tab: ParameterTable} and \ref{tab: NeuroTable} with $t_0$ varying from $0$ to $15$ hours.}
  \label{fig:NeuroModel015}
\end{figure}


\subsection{Sensitivity Analysis of Modified Model} \label{sec: Sensitivity Analysis of Modified Model}

We perform a similar sensitivity analysis on the modified model in \eqref{eq:ModifiedModel}, utilizing both the Morris and Sobol methods described in Section \ref{sec: Sensitivity Analysis} to quantify how uncertainty and variability in model output can be attributed to uncertainties in the inputs when adding in the neuroprotectant terms. As before, we consider as scalar response variables the ratio of the M1 to M2 microglia as in \eqref{eq:ratio} and the total activated microglia as in \eqref{eq:sum}.

Figure~\ref{fig:NPsens} shows the resulting parameter sensitivity rankings from the Morris elementary effects and Sobol sensitivity analysis when $t_0= 0$. Similar results hold when $t_0$ is varied across the admissible treatment time. In Figure~\ref{fig:ratioNPt0}, when considering the ratio of M1 to M2 microglia in \eqref{eq:ratio}, the switching rate of M2 to M1 microglia $s_{M2 \rightarrow M1}$ stands out as being the most sensitive, followed by the opposite switching rate $s_{M1 \rightarrow M2}$ and half-maximal concentration $K_P$ of pro-inflammatory cytokines. The parameters relating to the neuroprotectant terms (indexed 18-22) are much less sensitive in comparison. 

In Figure~\ref{fig:M1M2NPt0}, when considering the total activated microglia in \eqref{eq:sum}, the number of resting microglia $a$ remains the most sensitive parameter. The activation rate $k_{M2}$ of the M2 microglia becomes the second most sensitive parameter, followed by the natural mortality rate $\mu_{M2}$. Note that out of the neuroprotectant term parameters, the maximum value $L$ of the $N_2(t)$ term supporting M2 activation is the most sensitive.

\begin{figure}[t!] 
  \begin{subfigure}[b]{0.5\linewidth}
    \centering
    \includegraphics[width=1\linewidth]{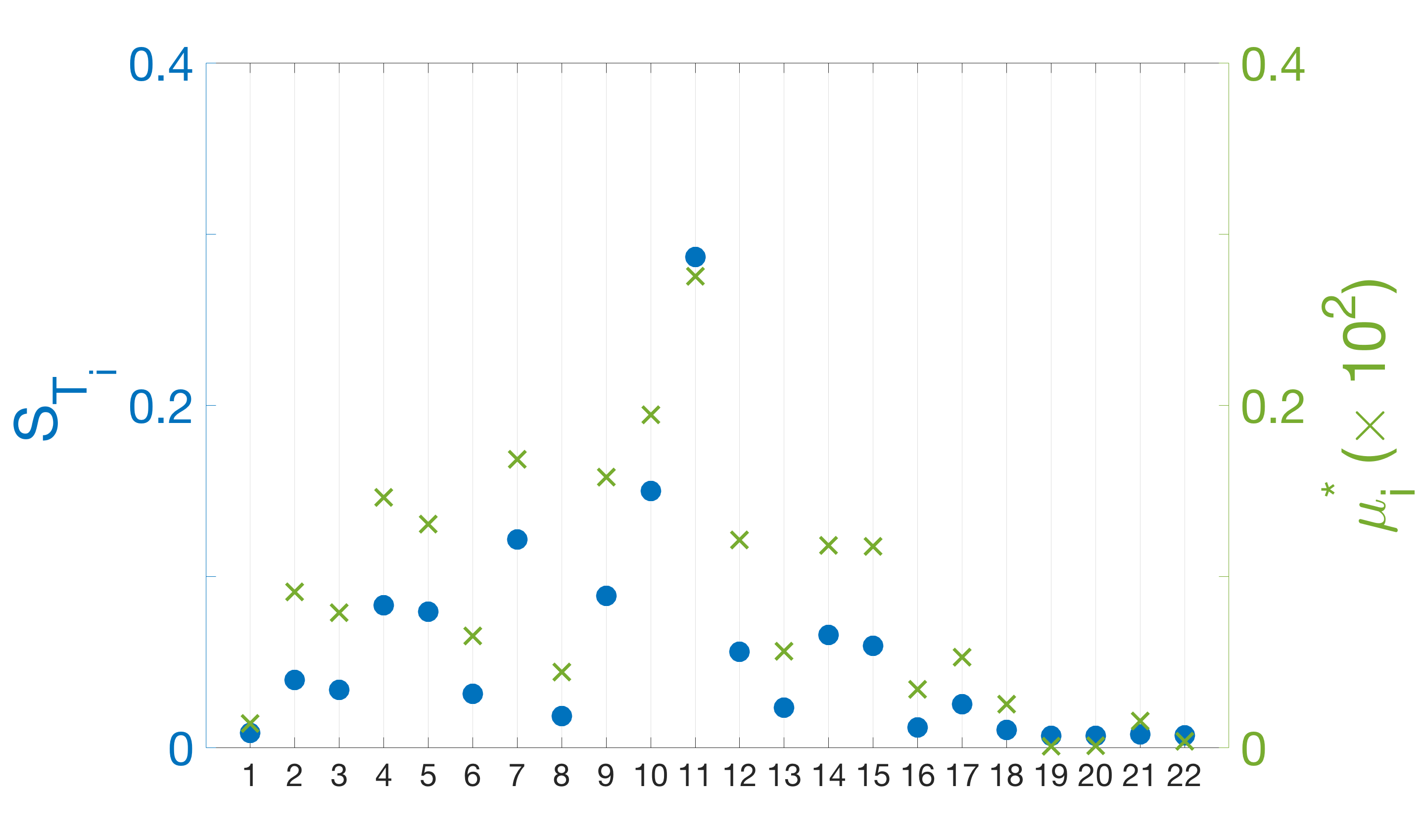} 
    \caption{$f(q) = \int_0^{72} M1(t;q)./M2(t;q) dt$ } 
     \label{fig:ratioNPt0}
    \vspace{4ex}
  \end{subfigure} 
  \begin{subfigure}[b]{0.5\linewidth} 
    \centering
    \includegraphics[width=1\linewidth]{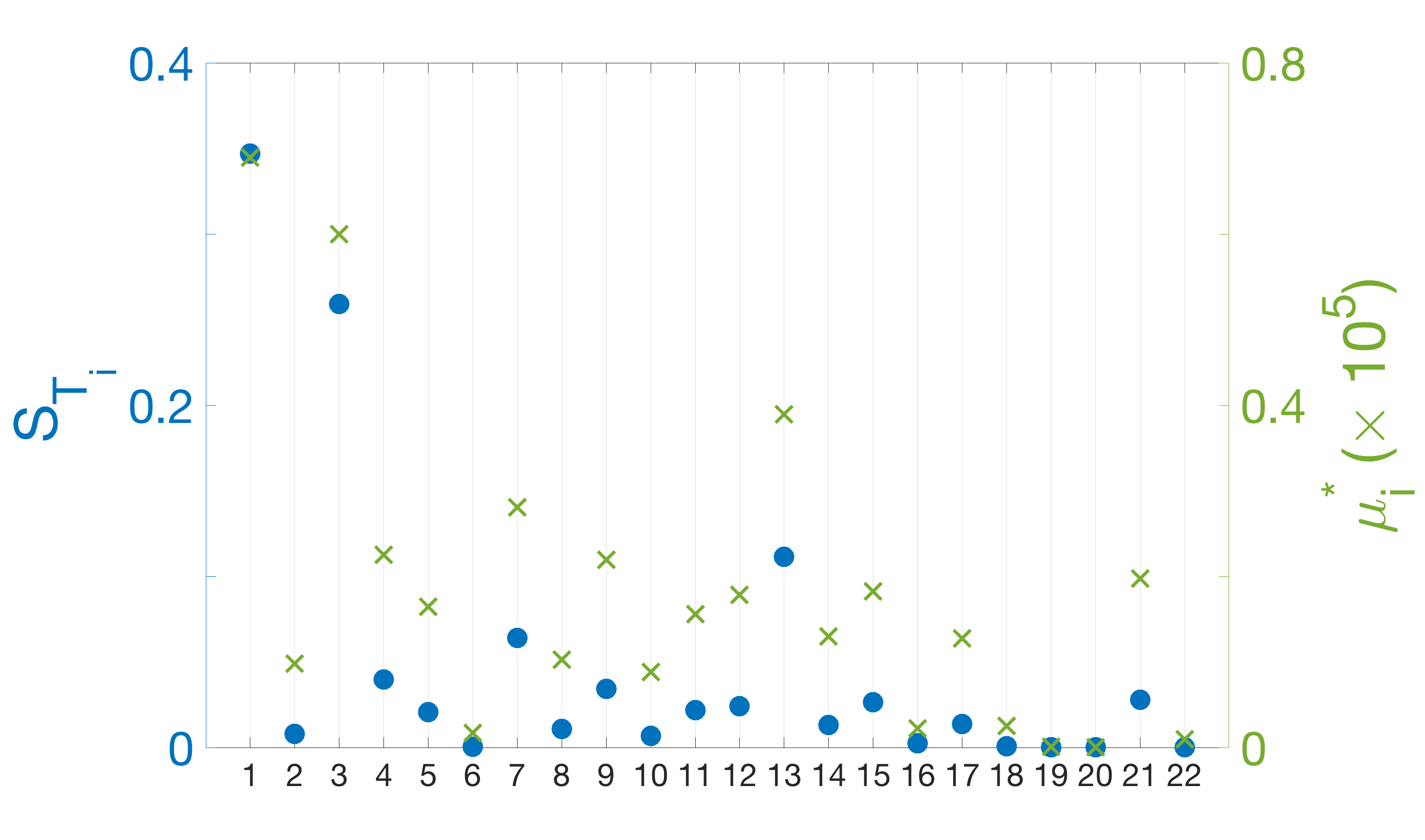} 
    \caption{$f(q) = \int_0^{72} (M1(t;q) + M2(t;q)) dt$ } 
    \label{fig:M1M2NPt0}
    \vspace{4ex}
  \end{subfigure} 
    \caption{Total-order sensitivity index $S_{T_i}$ (blue dot) and absolute mean $\mu_i^*$ (green x) for each parameter $q_i$ in Model \eqref{eq:ModifiedModel} with respect to the scalar response variable $f(q)$. Parameters are labeled using the indices $i$ ($i = 1, \dots, 22$) listed in Tables~\ref{tab: ParameterTable} and Table \ref{tab: NeuroTable} with $t_0 = 0$.}
  \label{fig:NPsens}
\end{figure}


\section{Discussion}  \label{sec: Discussion}

In this work, we develop a system of four coupled nonlinear differential equations describing the dynamics of activated microglia and cytokines during ischemic stroke. In particular, this model considers the switching of activated microglia between the M1 and M2 phenotypes (in both directions) and lumped compartments representing pro-inflammatory and anti-inflammatory cytokines. Inspired by possible neuroprotectant strategies, additional time-dependent terms are included to aid in the activation of beneficial M2 microglia and inhibit the activation of detrimental M1 microglia.

Simulations using nominal parameter values show that the model captures experimentally observed behavior of M1 and M2 microglial cells and cytokines post ischemic stroke. Numerical results further emphasize the importance of bidirectional switching between microglia phenotypes, in particular when considering the ratio of M1 to M2 microglia. Global sensitivity analysis results indicate that the parameters relating to phenotype switching in both directions are the two most influential parameters in the absence of terms to suppress M1 microglia and bolster M2 microglia. In the presence of these terms, the switching from M2 to M1 stands out as being the most sensitive parameter. These results indicate that the rate of switching between phenotypes in both directions is influential on the overall ratio of M1 to M2 microglia in the system.

By including terms to suppress the activation of M1 microglia and bolster the activation of M2 microglia, the model results in similar ratios of M1 to M2 microglia as observed in experimental studies for neuroprotectants. In particular, in numerical simulations using nominal parameter values, the modified model with neuroprotectant-inspired terms maintains a ratio of M1 to M2 cells below one for around 62 to 68 hours depending on the onset time. This is a significant extension of M2 cell dominance over results using the baseline model -- in the absence of neuroprotectant-based terms, the ratio remained under one for only the first 17 hours. Further, early onset time of the neuroprotectant terms leads to a decreased minimum ratio of M1 to M2 microglia, which suggests possible early reduction in the detrimental effects of neuroinflammation by maintaining a larger amount of M2 cells for a longer time period. 

When considering the total amount of activated microglia in the system, global sensitivity results intuitively show that model parameters relating to the M1 and M2 microglial cells are the most sensitive. These include the number of resting microglia and the activation and mortality rates of the M1 and M2 cells, while model sensitivity with respect to phenotype switching and cytokine production is negligible. Similar results are seen in both the absence and presence of the neuroprotectant terms, however the parameters relating to M2 become more sensitive than those for M1 when neuroprotective terms are included.  

When instead considering the ratio of M1 to M2 microglia, parameters involving phenotype switching and cytokines arise as being the most sensitive.  In particular, in the absence of the neuroprotectant terms, parameters relating to the half-maximal concentrations of anti-inflammatory and pro-inflammatory cytokines follow the switching parameters as the most sensitive, while the number of resting microglia is the least sensitive. Similar sensitivity rankings hold in the presence of the neuroprotectant terms, while the parameters relating to the neuroprotectant-inspired functions are notably some of the least sensitive parameters.

Since neuroprotectant strategies aim to decrease the ratio of M1 to M2 microglial cells while not necessarily altering the total number of activated microglia, it is of interest to further study model terms and parameters relating to the switching between microglia phenotypes, as well as those relating to cytokines. The proposed model can be extended to include separate compartments to account for interactions between microglia and specific pro-inflammatory cytokines (e.g., TNF-$\alpha$) and anti-inflammatory cytokines (e.g., IL-4, IL-10). 

Future work will be performed to estimate model parameters based on experimental data measuring the ratio of M1 to M2 microglia in the absence and presence of neuroprotectant treatment, including the use of time-varying parameter estimation techniques to estimate the $N_1(t)$ and $N_2(t)$ terms for specific neuroprotectants without assuming the piecewise sigmoidal forms. Further, compartments for additional intracellular components (including neurons, astrocytes, and macrophages) will be included to study their interactions with the microglial cells and effects on the ratio of M1 to M2 microglia. Additional terms may be included to model existing thrombolysis drug treatment and explore computational simulation of combination strategies with tPA and novel neuroprotectants.

\section{Conclusion} \label{sec: Conclusion}

Neural inflammation is a natural response following the onset of ischemic stroke, propagated by the activation of microglia. Resting microglia can become classically activated into the M1 phenotype and produce detrimental substances or alternatively activated into the M2 phenotype and produce beneficial substances. In this study, we formulate a nonlinear system of differential equations to model the interactions between the two microglia phenotypes and the cytokines that each phenotype produces during inflammatory response. Additionally, we include terms suppressing the activation of M1 microglia and bolstering the production of M2 microglia to simulate possible neuroprotectant strategies. Model simulations and global sensitivity analysis results highlight the significance of bidirectional microglia phenotype switching on the ratio of M1 to M2 microglia, in both the absence and presence of neuroprotectant-inspired terms. Simulation results further demonstrate that early onset of terms to inhibit M1 activation and support M2 activation leads to a decreased minimum ratio of M1 to M2 microglia and allows the M2 microglia to dominate the number of M1 microglia for a longer time window. 


\section*{Acknowledgments}

This work was partially supported by WPI's Presidential Fellowship (S. Amato) and by the National Science Foundation under grant number NSF/DMS-1819203 (A. Arnold).  \\

\vspace{.2cm}

\noindent \textbf{Declaration of Competing Interest:} None.\\





\end{document}